\documentclass[10pt,a4paper]{article}
\usepackage{graphicx}
\usepackage{tabularx}
\usepackage{multicol}
\usepackage{tabto}
\usepackage{url}
\usepackage{enumitem}
\usepackage{textcomp}
\usepackage{eurosym}
\usepackage{amsmath}
\usepackage{MnSymbol}
\usepackage{wasysym}
\usepackage{xcolor}
\usepackage[square,numbers]{natbib}
\usepackage[labelfont=bf]{caption}
\usepackage[yyyymmdd]{datetime}

\usepackage{flowchart}
\usetikzlibrary{shapes,arrows.meta,chains,external}

\def\ccol{0.7em}

\def\x{1}
\def\widefigure{}

\def\hackspace{-12pt}

\newcolumntype{L}[1]{>{\raggedright\arraybackslash}p{#1}}
\newcolumntype{C}[1]{>{\centering\arraybackslash}p{#1}}
\newcolumntype{R}[1]{>{\raggedleft\arraybackslash}p{#1}}

\newcommand{\eat}[1]{}

\newcommand{\cz}[1]{\textit{\textbf{#1}}}

\newcommand{\ts}{
    \tikzset{>={Latex[width=3mm,length=3mm]}}
    \tikzstyle{line} = [draw, ->, >=latex, ultra thick]
    \tikzstyle{circ} = [
        circle,
        align=center,
        text width=2em,
        text centered,
        inner sep=1mm,
        outer sep=1mm,
        minimum width=0cm,
        minimum height=0cm
    ]
    \tikzstyle{block} = [
        rectangle,
        draw,
        text width=7em,
        text centered,
        rounded corners,
        align=center,
        minimum width=\smbwd,
        minimum height=1cm
    ]
    \tikzstyle{fdecision} = [
        diamond,
        draw,
        text width=6.5em,
        text centered,
        align=center,
        aspect=1.5,
        minimum height=0.5cm,
        minimum width=\smbwd,
    ]
    \tikzstyle{parallelogram} = [
        trapezium,
        draw,
        trapezium left angle=70,
        trapezium right angle=-70,
        text width=0.5em,
        text centered,
        align=center,
        minimum width=\smbwd,
        minimum height=1cm
    ]
    \tikzstyle{box} = [draw, rectangle, thick, fill=gray!30]
    \tikzstyle{boxy} = [
        draw,
        rectangle,
        thick,
        text centered,
        align=center,
        minimum width=4cm,
        minimum height=4cm
    ]
    \tikzstyle{noshape} = [text width=5em, text centered, minimum height=5em]
    \tikzstyle{nos} = [
        rectangle,
        text width=6em,
        text centered,
        align=center,
        minimum width=4cm,
        minimum height=1cm
    ]
}

\begin{document}

\bibliographystyle{plain}
\thispagestyle{empty}

\begin{center}
\large{\bf A Digital Currency Architecture for Privacy and Owner-Custodianship}\\
\end{center}
\vspace{0.4em}
\begin{center}
\begin{minipage}[t][][t]{0.32\linewidth}
\begin{center}
\large{\bf Geoffrey Goodell}\\
\vspace{0.2em}
{\texttt{g.goodell@ucl.ac.uk}}
\end{center}
\end{minipage}
\begin{minipage}[t][][t]{0.32\linewidth}
\begin{center}
\large{\bf Hazem Danny Al-Nakib}\\
\vspace{0.2em}
{\texttt{h.nakib@cs.ucl.ac.uk}}
\end{center}
\end{minipage}
\begin{minipage}[t][][t]{0.32\linewidth}
\begin{center}
\large{\bf Paolo Tasca}\\
\vspace{0.2em}
{\texttt{p.tasca@ucl.ac.uk}}
\end{center}
\end{minipage}
\end{center}
\begin{center}
{\large University College London}\\
\vspace{0.4em}
{\large Centre for Blockchain Technologies}
\end{center}
\begin{center}
{\textit{This Version: \today}}\\
\end{center}

\abstract{
In recent years, electronic retail payment mechanisms, especially e-commerce
and card payments at the point of sale, have increasingly replaced cash in many
developed countries.  As a result, societies are losing a critical public retail
payment option, and retail consumers are losing important rights associated
with using cash.  To address this concern, we propose an approach to digital
currency that would allow people without banking relationships to transact
electronically and privately, including both e-commerce purchases and
point-of-sale purchases that are required to be cashless.  Our proposal
introduces a government-backed, privately-operated digital currency
infrastructure to ensure that every transaction is registered by a bank or
money services business, and it relies upon non-custodial wallets backed by
privacy-enhancing technology such as blind signatures or zero-knowledge proofs
to ensure that transaction counterparties are not revealed.  Our approach to
digital currency can also facilitate more efficient and transparent clearing,
settlement, and management of systemic risk.  We argue that our system can
restore and preserve the salient features of cash, including privacy,
owner-custodianship, fungibility, and accessibility, while also preserving
fractional reserve banking and the existing two-tiered banking system.  We also
show that it is possible to introduce regulation of digital currency
transactions involving non-custodial wallets that unconditionally protect the
privacy of end-users.
}

\section{Introduction}

As a form of money, cash offers many benefits to its owners. Cash is possessed
directly by its owners, allowing them to transact privately, without fear of
being profiled, discriminated against, or blocked, while knowing that their
money is as good as everyone else's. Cash is an obligation of the central
bank, and there is no intermediary that can default or abuse its trusted
position as custodian. In contrast, bank deposits are intrinsically custodial
and require their owners to trade the advantages of cash for a more limited set
of rights. In a rush to eliminate cash, societies risk imposing this choice
upon everyone. In this article, we ask whether digital currencies can offer
the benefits of cash in the age of electronic payments.

Mancini-Griffoli and his co-authors argue that anonymity is a salient feature
of cash, that privacy of transactions is essential, and that the specific
design features of central bank digital currency (CBDC) could have a
significant impact on financial integrity~\cite{mancini2018}. Our proposal
provides a solution with the flexibility to accommodate the widely-acknowledged
requirements and goals of CBDC and which is more akin to cash. Specifically,
it delivers a measure of control by restricting peer-to-peer transactions.
However, it does not offer the near-total degree of control that seems to be
taken as a requirement in some designs~\cite{auer2020}, and instead its retail
applications are exposed to a corresponding limitation to their scalability,
but not one that cannot be overcome by introducing additional control, in
limited contexts, outside the operating plane of the ledger.

We propose an approach that allows regulators to finely tune their choice of
trade-offs in a trilemma of scalability, control, and privacy. For example, it
might require that certain (or all) businesses cannot accept payments larger
than a certain size without collecting or reporting additional information that
limits privacy, or it might require that some individuals or non-financial
businesses have a larger or smaller cap on the volume of their withdrawals into
non-custodial wallets. To draw an analogy, it operates like an automated
conveyor belt holding keys that are trying to meet a lock, and if they are the
right fit, as determined either at large or on a case-by-case basis, then the
transactions take place in an automated way. For the avoidance of doubt, such
automation can include so-called ``embedded transactions'' wherein payments can
be seamlessly integrated into the transaction without independent mechanisms or
reconciliation.

The rest of this article is organised as follows. The next section provides
essential background on digital currencies, distributed ledger systems, and
privacy-enhancing technologies. Section~\ref{s:design} introduces our approach
to digital currency that offers strong privacy and owner-custodianship while
also facilitating essential regulatory oversight. Section~\ref{s:analysis}
provides an analysis of the salient features of our design. The final two
sections offer our recommendations and conclusion, respectively.

\section{Background}
\label{s:background}

In this section, we contextualise and offer motivation for our proposal for
CBDC from four different angles: first, by identifying the need for a cash-like
public mechanism for retail electronic payments; second, by acknowledging the
current zeitgeist in which CBDC is being considered by central banks and
institutions; third, by identifying distributed ledgers and privacy-enhancing
cryptographic techniques as essential enabling technologies for our proposal;
and, finally, by exploring the system governance requirements of our proposal in
the context of existing systems that share some common characteristics.

\subsection{Cash for the Digital Age}
\label{ss:cash}

Although retail digital currency transactions are currently perceived as
something of a niche market, reason exists to believe that the scope and set of
use cases for such transactions will expand in the decades ahead.\footnote{The
text of Section~\ref{ss:cash}, with the exception of the last two paragraphs,
also appears in a response to a recent consultation by the US Financial Crimes
Enforcement Network~\cite{goodell2021}.} One important reason relates to the
secular decline in the use of cash in much of the developed world. Indeed,
many retailers have come to conclude that accepting cash is optional, and for
this reason legislation to compel retailers to accept cash exists in many
jurisdictions around the world, including Denmark, Norway, China, and several
US states~\cite{access,sadeghi}. However, such legislative protections might
not be enough to sustain cash as a viable payment option. As retail
transactions increasingly take place electronically, the variable revenues
associated with operating cash infrastructure fall relative to the fixed costs,
and the marginal cost of handling cash increases. This logic applies without
distinction to retail users, including both customers and vendors, as well as
banks and operators of ATM networks. In the UK, ATM networks and bank branches
that facilitate the circulation of cash are facing pressure that has led to a
downward spiral in cash services~\cite{tischer}.

Cash specifically confers certain important advantages to its bearers that
modern retail payment infrastructure does not, including but not limited to:

\begin{itemize}

\item \cz{Owner-custodianship.} The absence of a custodian means that the
bearer cannot be blocked by the custodian from making a remittance or charged
differentially by the custodian on the basis of the counterparty to a
transaction. Self-determination is an essential feature of ownership, and a
critical prerequisite to ownership is the ability to withdraw and use cash in a
multitude of transactions without a custodian.

\item \cz{True fungibility.} Because cash does not require any particular
identification or imply any particular relationship with a financial
institution, users of cash know that their money is exactly as valuable as
anyone else's.\footnote{In some parts of the world, tattered banknotes might be
less valuable than pristine ones, and in such circumstances even cash might not
be entirely fungible.} Absent this property, counterparties to a transaction
would be able to discriminate on the basis of the identity of the bearer or the
custodian, and the same amount of money would have a different value in the
hands of different people.

\item \cz{Privacy by design.} It is no secret that retail payments leave
behind a data trail that can be used to construct a detailed picture of an
individual's personal lives, including travel, financial circumstances,
relationships, and much more. The fact that electronic payments can be used
for surveillance and population control has been known for many
decades~\cite{armer1968,armer1975}. We further note that data protection,
which relates to the access and use of private information once collected, is
not the same as privacy by design, wherein users of a technology do not reveal
private information in the first instance. The importance of favouring privacy
by design to data protection is well-understood~\cite{nissenbaum2017}, and the
continued inability of governments and corporations to prevent unauthorised
access, both by (other) government authorities and by malicious adversaries,
underscores a greater need for private information to not be
collected~\cite{rychwalska}. This argument has also been specifically
elaborated in the context of value-exchange systems~\cite{goodell20a}.

\end{itemize}

Non-custodial wallets offer a way to preserve cash-like characteristics in
digital transactions, and we have argued that the popularity of
cryptocurrencies largely follows from the pursuit of privately held digital
cash~\cite{goodell2019}. We suggest that non-custodial wallets should offer to
their users the same affordances as cash. Consequently, they are essential to
individual privacy and human rights. There is no reason to assume that the
increasing preponderance of online and digital transactions must present an
opportunity to expand the scope for surveillance and control over individual
persons by monitoring or restricting what they do with their money.

In the context of CBDC, non-custodial wallets offer a direct \textit{economic}
relationship, but not a direct \textit{technical} relationship, between retail
CBDC users and the central bank. By this, we mean that CBDC tokens would
constitute a liability of the central bank.  This is consistent with the
current two-tiered banking system.  We do not mean to suggest that retail CBDC
users would have accounts with the central bank or that they would interface
with the central bank directly.

\subsection{CBDC and Private-Sector Banks}
\label{ss:banks}

In May 2020, Yves Mersch, Vice-Chair of the Supervisory Board and Member of the
Executive Board of the European Central Bank, stressed the importance of the
role of the private sector in operating a network for payments:

\begin{quote}
``[D]isintermediation would be economically inefficient and legally untenable.
The EU Treaty provides for the ECB to operate in an open market economy,
essentially reflecting a policy choice in favour of decentralised market
decisions on the optimal allocation of resources. Historical cases of
economy-wide resource allocation by central banks are hardly models of
efficiency or good service. Furthermore, a retail CBDC would create a
disproportionate concentration of power in the central
bank.''~\cite{mersch2020}
\end{quote}

A few months before Mersch's speech, Tao Zhang, Deputy Managing Director of the
International Monetary Fund (IMF), also offered his opinion on the current set
of proposals for CBDC, which he said ``imply costs and risks to the central
bank''~\cite{zhang2020}. We argue that his conclusions follow from the
proposals that have been elaborated so far by central banks, which have
generally involved a central ledger operated by the central bank
itself~\cite{mas2018,boe2020}. We suggest that such proposals have been
designed neither to be holistic nor to complement the current model of
payments, settlement, and clearing that exists today. In contrast, our
approach specifically avoids the costs and risks identified by Mersch and
Zhang, which we characterise more specifically in Section~\ref{ss:dlt}, and is
broadly complementary to the current system.

Zhang also introduced the idea of a ``synthetic CBDC'' consisting of tokens
issued by private-sector banks~\cite{zhang2020}. We argue that the desirable
qualities that Zhang ascribes to synthetic CBDC apply to our proposed solution
as well, except that our proposed solution still allows for ``real'' CBDC
whilst the infrastructure would be operated by private-sector \textit{money
services businesses} (MSBs), including but not limited to banks, and, for our
purposes, comprise both traditional commercial banks and financial institutions
as well as new entities that would only have central bank reserves as their
assets and whose liabilities would, in turn, only be deposits. This is an
important distinction, and although Zhang provides no specific description of
the technical features of synthetic CBDC, we assume that it would not involve a
distributed ledger and that it would not be possible to have private
transactions, since the private-sector banks would have visibility into the
operation and ownership of their own tokens.

Nevertheless, an effective retail CBDC does not necessitate disintermediation
of the banking sector. The CBDC that we envision would have more in common
with physical cash than with bank deposits, and it would not substitute for
bank deposits. It would not be eligible for rehypothecation and would not pay
interest to its bearers, at least not in the traditional sense. We view retail
CBDC principally as a technology to facilitate payments and consumer
transactions. It is not simply a more scalable version of wholesale CBDC,
reflecting the fact that the requirements for retail and wholesale users of
money are not the same. Retail CBDC users would have the same reasons to
favour bank deposits over CBDC for their long-term investments for the same
reason that they favour bank deposits over cash for the same purpose; we
discuss this further in Section~\ref{ss:liquidity}. We also note that a
central bank would not be a valid substitute for commercial banks, which we
discuss further in Section~\ref{ss:comparison}.

\subsection{Architectural Considerations}

Another critical question is whether CBDC should be ``account-based'', by which
we mean that users would strictly interact with \textit{accounts} representing
long-lived relationships, or ``token-based'', by which we mean that CBDC would
exist independently of any particular relationship, as coins and bank notes do.
Accounts can represent relationships with a custodian or with the ledger system
itself, and not all digital currency designs are the same. For example,
although tokens in Bitcoin are explicitly designed to exist
independently~\cite{nakamoto}, tokens in Ethereum are explicitly designed to
exist within accounts~\cite{buterin}. The two architectures are not symmetric:
Although tokens in token-based systems can be held by custodians on behalf of
users, such an arrangement is optional, whereas accounts are intrinsically
designed to represent a persistent relationship.

MSBs do not necessarily perform all of the functions of banks, such as lending
credit. Moreover, in our design, we envisage full convertibility at par across
CBDC, bank deposits, bank notes, and (for authorised MSBs) reserves, both to
ease its introduction and to not interfere with the fungibility and general
composition of the monetary base. To whatever extent this involves limitations
or the introduction of frictions will be a matter of policy. Yet, in
principle, at-par convertibility for cash and bank deposits as the default is a
practical and design necessity. Issuing and introducing CBDC enables a new
policy tool in adjusting the incentives to hold or spend the CBDC through its
various features but also to balance the possible flight from bank
deposits~\cite{10}, for which we do not see CBDC as a general substitute.

\subsubsection{Distributed Ledger Technology}
\label{ss:dlt}

\noindent Distributed Ledger Technology (DLT) offers a way to share
responsibility for rulemaking among a set of peers. A \textit{distributed
ledger} is ``a ledger that is shared across a set of DLT nodes [peers] and
synchronized between the DLT nodes using a consensus
mechanism''~\cite{iso22739}. Although it is theoretically possible to build
public digital currency infrastructure, even privacy-preserving digital
currency infrastructure, using centralised technology~\cite{chaum2021}, we
argue that the salient features of a distributed ledger, including without
limitation community \textit{consensus} and
\textit{immutability}~\cite{iso22739}, are necessary for the infrastructure to
succeed in practice. This should not be interpreted to mean that the
infrastructure must provide for or allow peer-to-peer transactions among users.
This should be interpreted to mean that the system must be operated by a
community, not some privileged arbiter, and that the consensus view of the
truth about which transactions have taken place should reflect the agreement of
this community. In particular, we rely upon DLT to marshal consensus among
independent actors so that substantially all of the community must agree before
a new entry is added to the ledger or before the rules governing the operation
of the ledger are changed.

In the context of digital currency, DLT would provide transparency to the
operation and rules of the system by restricting (at a technical level) what
any single actor, including the central bank, as well as government regulators,
can decide unilterally. Such transparency complements and does not substitute
for regulatory oversight.

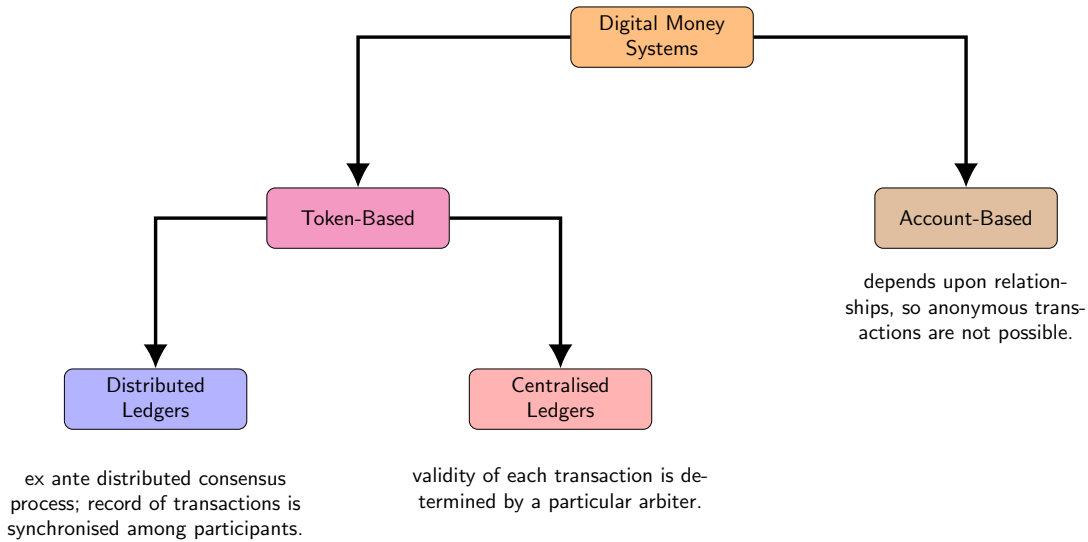
\begin{figure}
\widefigure
\begin{center}
\scalebox{0.8}{\hspace{\hackspace}\begin{tikzpicture}[
  >=latex,
  node distance=3cm,
  font={\sf},
  auto
]
\ts
\tikzset{>={Latex[width=4mm,length=4mm]}}
\tikzstyle{noshape} = [
  rectangle,
  text width=14em,
  text centered,
  align=center,
  minimum width=8cm,
  minimum height=1cm
]
\def\smbwd{3cm}

\node (block1) at (0,0) [block, fill=orange!50] {Digital Money Systems};
\node (block2) at (-5,-3) [block, fill=magenta!50] {Token-Based};
\node (block3) at (5,-3) [block, fill=brown!50] {Account-Based};
\node (block4) at (-8.33,-6) [block, fill=blue!30] {Distributed Ledgers};
\node (block5) at (-1.67,-6) [block, fill=red!30] {Centralised Ledgers};

\node (desc2) at (5, -4.5) [noshape] {depends upon relationships, so
anonymous transactions are not possible.};

\node (desc4) at (-8.33, -7.75) [noshape] {ex ante distributed consensus process;
record of transactions is synchronised among participants.};

\node (desc5) at (-1.67, -7.5) [noshape] {validity of each transaction is
determined by a particular arbiter.};

\draw[->, ultra thick] (block1) -| (block2);
\draw[->, ultra thick] (block1) -| (block3);
\draw[->, ultra thick] (block2) -| (block4);
\draw[->, ultra thick] (block2) -| (block5);

\end{tikzpicture}}

\caption{\cz{Taxonomy of Digital Money Systems.}}

\label{f:taxonomy}
\end{center}
\end{figure}

Figure~\ref{f:taxonomy} shows a taxonomy of digital money systems. Digital
money systems include CBDC. The first question to ask is whether we need a
system based on tokens rather than a system based on accounts. There are
several benefits to using a token-based system, including substantially
reducing the overhead associated with pairwise reconciliation and regulatory
reporting. Most importantly, however, any system based upon accounts cannot
offer privacy, since its design would necessarily require resolvable account
identifiers that can ultimately be used to determine both counterparties to any
transaction. Therefore, we must recognise that preservation of a token-based
medium of exchange is necessary to the public interest, increases welfare, and
maintains the critical nature of cash while providing to central banks and
governments the assurance and risk assessment tools that are afforded to
digital payment infrastructure platforms.

\ifnum\x=0
  \end{paracol}
\fi
\ifnum\x=0
  \begin{paracol}{2}
  \linenumbers
  \switchcolumn
\fi

\subsubsection{Privacy by Design}
\label{ss:privacy}

We suggest as a design requirement that retail CBDC users must have the right
to privacy, with respect to not only asset custodians and other corporate
actors but also the state. Law enforcement can ask custodians to carry out
legitimate law-enforcement capabilities. However, it is too easy to assume
that all of the information about a transaction should be available to law
enforcement (or others) for their perusal upon request, and it has become an
accepted practice for governments to leverage relationships between individuals
and private-sector businesses to extract such information about their
transactions.

Fortunately, it is possible to regulate financial transactions without
collecting data that could be used to profile the behaviour of individual
persons. Features that can be applied on a case-by-case basis, such as limits
to the size of withdrawals to anonymous destinations or limits to the size of
remittances into accounts from private sources, can be implemented externally
to the core architecture and managed by policy.

We do not envision privacy as something that can be bolted on to a
fully-traceable system (for example, with ``anonymity
vouchers''~\cite{r3-cbdc,dgen}) or that can depend upon the security or
protection offered by some third party. Were a CBDC designed not to provide
certain qualities of privacy, some users, including those concerned about the
risk of profiling or discrimination, would have reason to continue using less
well-regulated methods to transact, including cash~\cite{agur2019}. Moreover,
there is demand for semi-anonymous means of payment~\cite{20}, as well as for a
variety of different instruments that can be used for payment, and, due to
heterogeneity in the preferences of households, the use of a CBDC has immediate
social value~\cite{21}.

In the same statement mentioned in Section~\ref{ss:banks}, Mersch also
specifically acknowledged the importance and significance of preserving
privacy, suggesting that an attempt to reduce the privacy of payments would
``inevitably raise social, political and legal issues''~\cite{mersch2020}.

This acknowledgement is important for three reasons. First, no digital
currency, token-based or otherwise, would guarantee perfect anonymity in
practice: consider the potential for timing attacks, software bugs, and
limitations to operational security. Even bank notes do not achieve perfect
anonymity: their serial numbers offer a possibility wherein individual notes
can be tracked or marked, although to our knowledge such methods for
surveillance are imperfect and seldom used. Nevertheless, we must consider the
implications of systems that attempt to force users into payment systems with
different anonymity properties and trade-offs in general. Second, we have an
opportunity to demonstrate a system that can achieve and deliver a measure of
true privacy, in contrast to problematic assumptions, such as the idea that the
system must accommodate exceptional access or that privacy is not the starting
point but rather something that should be protected by an
authority~\cite{benaloh2018}. Such a system, an example of which we describe in
Section~\ref{s:design}, would constitute an improvement over both the various
government-backed digital currency systems that have been proposed to date
(which are institutionally supportable but not private), as well as the various
``outside solutions'' involving \textit{permissionless} ledgers that are used
in cryptocurrencies, such as Zcash and Monero (which are potentially private
but not institutionally supportable). Third, privacy is sufficiently important
that we should not rush headlong into creating infrastructure, or allowing
infrastructure to be created, that might forcibly undermine it. In contrast to
\textit{data protection}, which is about preventing unauthorised use of data
following its collection, \textit{privacy} is about preventing individuals (and
in some cases businesses) from revealing information about their (legitimate)
habits and behaviours in the first instance. Data protection is no substitute
for privacy by design~\cite{nissenbaum2017}. As an architectural property,
therefore, privacy is a fundamental design feature that cannot be ``granted''
or ``guaranteed'' by some authority.

In principle, it should be possible to accommodate privacy by design with a
regulatory approach that intrinsically protects the rights of retail CBDC
users.\footnote{The text of the remainder of Section~\ref{ss:privacy} also
appears in a response to a recent consultation by the US Financial Crimes
Enforcement Network~\cite{goodell2021}.} To avoid infringing upon essential
privacy and human rights, specific regulatory and technical measures must be
taken to ensure:

\begin{itemize}

\item that non-custodial wallets must not be expected to carry persistent
identifying information, such as a unique identifier or address that would be
associated with multiple transactions,

\item that non-custodial wallets must not be expected to reveal information,
including keys or addresses associated with previous or subsequent
transactions, that can be used to identify their bearers, owners, or sources of
funds;

\item that the obligation to identify the counterparties to a transaction can
only be imposed at the time of a transaction; and

\item that the process for providing information to the requesting banks or
MSBs for the purposes of recordkeeping or reporting must not involve the
non-custodial wallet itself and would be carried out only with the consent of
both counterparties.

\end{itemize}

It can only be possible for ordinary users of non-custodial wallets
to have confidence that their routine activities will not be profiled if the
relevant thresholds are sufficiently high and circumstances are sufficiently
rare for which counterparty information is requested for recordkeeping or
reporting. Such requests must involve the explicit consent of the owner or
bearer of the digital tokens on each separate occasion, must not be routine for
ordinary persons carrying out ordinary activities, and must not require a
non-custodial wallet or any other personal device to reveal any information
identifying its owner or bearer.

\subsubsection{Privacy-Enhancing Cryptographic Techniques}
\label{ss:pets}

To achieve privacy by design, we consider applying privacy-enhancing technology
of the sort used by privacy-enabling cryptocurrencies, such as Zcash and Monero.
There are at least three possible approaches:

\begin{enumerate}

\item \cz{Stealth addresses, Pedersen commitments, and ring signatures.}
Stealth addresses, which obscure public keys by deriving them separately from
private keys~\cite{courtois2017}, deliver privacy protection to the receiver of
value~\cite{iso-tr-23244}. Pedersen commitments, which obscure the amounts
transacted to anyone other than the transacting
parties~\cite{pedersen1991,vanwirdum2016}, remove transaction metadata from the
ledger records~\cite{iso-tr-23244}. Ring signatures, which allow signed
messages to be attributable to ``a set of possible signers without revealing
which member actually produced the signature''~\cite{rivest2001}, deliver
privacy protection to the sender of value~\cite{iso-tr-23244}.

\item \cz{Zero-knowledge proofs (ZKP).} ZKP ``allow[s] one party to prove to
another party that a statement is true without revealing any information apart
from the fact that the statement is true''~\cite{iso-tr-23244} and can
potentially be used to protect all of the transaction
metadata~\cite{iso-tr-23244}. Non-interactive approaches to ZKP such as
ZK-STARKs deliver significant performance advantages over their interactive
alternatives~\cite{ben-sasson2018}, and based upon their measured
performance~\cite{ben-sasson2018,guan2019,zcash-sapling}, we anticipate that
such operations should be fast enough to suffice for point-of-sale or
e-commerce transactions, although stronger evidence to establish this fact
would be useful.

\item \cz{Blind signatures or blind ring signatures.} Because our proposed
architecture is not peer-to-peer with respect to its users, we believe that it
is also possible to combine a blind signature approach similar to the one
suggested by Chaum~\cite{chaum2021}. Using this method would require either
that recipients of (spent) tokens would immediately redeem them with the issuer
or, equivalently, that the payer would anonymously instruct the issuer to
deposit funds into an account designated for the recipient. With blind ring
signatures, the issuer could be a set of actors that work independently and
either deposit or withdraw funds into a shared account whenever they,
respectively, issue or redeem tokens, rather than a single
actor~\cite{ghadafi2013}.

\end{enumerate}

\subsection{System Governance}
\label{ss:governance}

Because privacy-enhancing technologies require vigilance~\cite{zimmermann1991},
MSBs and the broader community must commit to maintain, audit, challenge, and
improve the technology underpinning the privacy features of this design as part
of an ongoing effort~\cite{goodell2019}. Such maintenance implies establishing
a process for security updates, as well as updates to accommodate new technology
and features as needed. The transparency afforded by the use of DLT can
provide the basis by which the broader community can observe and analyse the
operation of the system, including any changes to its regular functioning, to
ensure that transacting parties remain protected against technologically
sophisticated adversaries with an interest in de-anonymising the CBDC users for
the purpose of profiling them.

Ultimately, whoever controls the code that the system relies upon to operate,
controls the operation of the system. By analogy, consider the role of
developer communities in handling ledger-related disputes in cryptocurrency
communities~\cite{wong2016}. For this reason, a centralised developer community
could certainly negate the benefit of a decentralised ledger. This implies that
each independent participant in the system should establish its own rigorous
procedure for accepting changes to the code, most likely including internal
code review and security analysis, whether or not participants share the same
code base, and it might be necessary for this process to be subject to public
oversight, as well. Such procedures for internal and external oversight should
involve a broad security community with diverse allegiances, and in particular,
care must be taken to ensure that it will be possible to make timely changes to
address emerging problems (including but not limited to shutdowns and partial
shutdowns) while protecting both users and system operators from the
possibility that backdoors or other vulnerabilities might be introduced in
haste. This is no simple task, although the work of the security community in
free software projects, such as Debian~\cite{debian-security} demonstrate that
the combination of deep oversight and timely changes is possible.

Additionally, we note that regulators can work with the private sector to
develop rules and enforce compliance procedures. For example, consider the
established procedures for the operation of trading networks, such as the
National Market System in the United States~\cite{nms-changes}. In 2005, the US
Securities and Exchange Commission instituted Rule 611, which specifies that
all FINRA member firms must publish and subscribe to a real-time feed listing
the best bids and offers for every listed security from among all of the
exchanges in the system, and that any exchange that receives a marketable order
must route it to the exchange that would execute the order at the best price.
Thus, FINRA member firms were required to implement advanced technology to
ensure that all marketable orders are routed correctly~\cite{rule611}.

Such examples demonstrate that regulators can specify rules with significant
implications for technology, that technology can be developed in support of
such rules by private-sector actors, and that changes to the rules can be
undertaken in a co-regulatory context, with formal proposals by regulators. In
our view, it is better not to allow prejudices about the technical
sophistication of government actors to limit our ambitions for public systems.

From the standpoint of CBDC, platform governance and decision-making
predominantly relates to authenticating and thereby allowing transactions. We
contend that the infrastructure underlying our proposal would be overseen by
the public sector but can be exclusively operated by the private sector. We
envisage that there should be no fewer than five MSBs for a pilot, and no fewer
than about twenty MSBs for robust operation. The approval of transactions takes
place through consensus across the infrastructure operators of the platform.
However, the ability to formally become an infrastructure operator and MSB
pro tanto requires the approval of the local regulator, however it is
regulated. We assume in this context the central bank is responsible for
overseeing clearing and settlement activities.\footnote{For example, in the
case of the United Kingdom it may be through joint oversight between the
Prudential Regulatory Authority (PRA) and the Financial Conduct Authority
(FCA) for matters related to conduct.}

\section{Our Proposal}
\label{s:design}

The core of our proposed design is based upon an article by Goodell and
Aste~\cite{goodell2019}, which describes two approaches to facilitate
institutional support for digital currency. We build upon on the second
approach, \textit{institutionally-mediated private value exchange}, which is
designed to be operated wholly by regulated institutions and has the following
design features:

\begin{enumerate}

\item Provides a \textit{government-issued electronic token} that can be used
to exchange value without the need for pairwise account reconciliation.

\item Allows transaction infrastructure (payments, settlement, and clearing) to
be operated by \textit{independent, private actors}\footnote{Presumably, the
independent, private actors would participate in the activities of a
co-regulated authority, such as FINRA in the United States, or a quango, such
as FCA in the United Kingdom.} while allowing central banks to control monetary
policy and CBDC issuance, with control over the creation and destruction of
CBDC but not its distribution.

\item Protects the \textit{transaction metadata} linking individual CBDC users
to their transaction history by design, without relying upon trusted third
parties.

\item Affords regulators \textit{visibility} (but excluding counterparty
information) into every transaction, allowing for analysis of systemic risks.

\end{enumerate}

In this section, we describe how our proposed mechanism for digital currency
works at a system level, identifying the interfaces between the institutional
and technical aspects of the architecture.

\subsection{Assumptions}
\label{ss:assumptions}

We imagine that digital currency might be issued by a central bank as ``true''
\textit{central bank digital currency} (CBDC), although it might alternatively
be issued by the government, representing an obligation on a collateralised
collection of State assets, such as sovereign wealth or Treasury assets. In
either case, we note that, in many countries (including the UK), no single party
(including the central bank) has been assigned the responsibility to design,
maintain, and update the rules of the process by which financial remittances
are recorded and to adjudicate disputes concerning the veracity of financial
remittances. We also note that the responsibility to operate transaction
infrastructure and supervise payment systems is different from the
responsibility to create tokens and safeguard the value of State currency. In
many countries, systems for payments, clearing, and settlement are a
collaborative effort~\cite{bis2012,bis2012a}. A design that externalises
responsibility for the operation of a transaction infrastructure supporting
digital currency is not incompatible with the operational role of a central
bank in using digital currency to create money and implement monetary policy.

In particular, we question the argument that because the central bank has no
obvious incentive to abuse data, therefore, all users should be expected to
trust it with their payments data. The idea of furnishing authorities with
exceptional access to private data, including specifically the idea of dividing
access to private data among multiple authorities, has been
debunked~\cite{abelson2015}. In particular, an apparently disinterested actor
can quickly become an interested actor when it finds itself in possession of
something that is of interest to its influential neighbours. So, we might
reasonably trust a central bank with monetary policy but not with transaction
data.

Our approach to digital currency differs substantively from the vision proposed
by several central banks~\cite{mas2018,boe2020}. We argue that the purpose of
digital currency is to provide, in the retail context, a mechanism for
electronic payment that does not rely upon accounts, and in the wholesale
context, a means of settlement that is more robust and less operationally
burdensome than present approaches. It is not to create a substitute for bank
deposits, which would still be needed for economically important functions such
as fractional reserve banking, credit creation, and deposit insurance. Neither
is it a replacement for cash, which offers a variety of benefits including
financial inclusion, operational robustness, and the assurance that a
transaction will complete without action on the part of third parties. We
imagine that in practice, digital currency would be used primarily to
facilitate remittances that cannot be done using physical cash and that people
would not be more likely to be paid in digital currency in the future than they
would to be paid in cash today.

Nevertheless, we intend our proposed design to replicate some of the features
of cash. Specifically, we seek to achieve the following properties:

\begin{enumerate}

\item\cz{Resistance to mass surveillance.} Cash allows its bearers to transact
without fear that they will be profiled on the basis of their activities. In
Section~\ref{ss:fraud}, we shall explicitly demonstrate that our design is
unlikely to increase the risk of fraud or AML/KYC violations relative to the
current system by comparing our proposed system to cash. In fact, we suspect
that it will lead to the opposite effect, given the possibility for the use of
digital analysis tools in the cases of regulated activities wherein adherence
to certain specific compliance rules is required and analysis over regulated
institutions activities is helpful.

\item\cz{Transaction assurance.} Cash allows its bearers to know that a
potential transaction will succeed without depending upon a custodial or
third-party relationship that might block, delay, or require verification for a
transaction to take place.

\item\cz{Non-discrimination.} Cash allows its bearers to know that their money
is as good as everyone else's, and specifically that its value is not
determined by the characteristics of the bearer.

\end{enumerate}

To achieve these requirements, our approach must be ``token-based'', by which
we mean that retail users must be able to hold tokens representing value
outside of custodial relationships and that the tokens are not forcibly linked
to an address or identifier that can be used to identify the user or the user's
other tokens. Accounts can be used in conjunction with the token
infrastructure, although we specifically disagree with the argument offered by
Bordo and Levin that suggests that only accounts can pay interest and therefore
all CBDC should be held in accounts~\cite{16}. In particular, it is not
obvious that a CBDC system should pay interest to its bearers; we note that
cash does not (see Sections~\ref{ss:cash}
and~\ref{ss:retail}).\footnote{Separately, we believe that it is possible for a
ledger system to offer interest-like remuneration, with some important
limitations, directly to tokens themselves.} Specifically, the trust property
we seek is intrinsic to the token, in that we want retail users to trust the
token itself and not some particular set of account-granting institutions or
system operators. We also explicitly state: \textit{Trust cannot be
manufactured and must be earned}. More importantly, we do not create trust by
asking for it; we create trust by showing that it is not needed. The approach
that we describe in Section~\ref{s:design} addresses this requirement directly.

We imagine that many, but not necessarily all, ordinary people and businesses
would have bank accounts into which they would receive payments. These bank
accounts would sometimes earn interest made possible by the credit creation
activities of the bank. Banks would be able to exchange digital currency at
par for cash or central bank reserves and would not generally hold wallets
containing an equal amount of digital currency to match the size of their
deposits. In the case of CBDC, banks would also be able to directly exchange
the digital currency for central bank reserves. When an individual (or
business) asks to withdraw digital currency, the bank would furnish it, just as
it would furnish cash today. The bank might have a limited amount of digital
currency on hand just as it might have a limited amount of cash on hand to
satisfy such withdrawal requests, and there would be limits on the size and
rate of such withdrawals just as there would be limits on the size and rate of
withdrawals of cash. Once they have digital currency, individuals and
businesses could use it to make purchases or other payments, as an alternative
to account-based payment networks or bank transfers, and digital currency would
generally be received into wallets held by regulated MSBs, just as cash would
be.

\subsection{System Design Overview}

Our design for CBDC is based on the approach described as an
\textit{institutionally mediated private value exchange} by Goodell and
Aste~\cite{goodell2019}, which we elaborate here and further build upon. This
proposal uses DLT for payments, as motivated by reasons articulated in
Section~\ref{ss:dlt}.

We envision a \textit{permissioned} distributed ledger architecture wherein the
participants would be regulated MSBs. MSBs would include banks, other
financial businesses, such as foreign exchange services and wire transfer
services, as well as certain non-financial businesses, such as post
offices~\cite{bis2012}, as well. In contrast to \textit{permissionless} DLT
systems that require computationally expensive and resource-intensive
mechanisms, such as proof-of-work to achieve resistance to Sybil attacks, the
\textit{permissioned} DLT design would support efficient consensus mechanisms,
such as Practical Byzantine Fault Tolerance~\cite{castro1999}, with performance
that can be compared to popular payment networks. In particular, Ripple has
demonstrated that its network can reliably process 1500 transactions per
second~\cite{ripple2019}. Although the popular payment network operator Visa
asserts that its system can handle over 65,000~transactions per
second~\cite{visa2018}, its actual throughput is not more than 1700
transactions per second~\cite{visa2020}. For this reason, we anticipate that
it will be possible for a digital currency solution to achieve the necessary
throughput requirement without additional innovation. Additionally, although
the distributed ledger would require peer-to-peer communication among
participants, we anticipate that the resource consumption would be comparable
to the data centres that house securities exchanges or clearing networks, in
contrast to the data centres that house so-called ``mining pools'' for
permissionless DLT networks.

We assume that the only parties that could commit transactions to the ledger
and participate in consensus would be MSBs, which would be regulated entities.
The ledger entries would be available for all participants to see, and we
imagine that certain non-participants, such as regulators and law enforcement,
would receive updates from the MSBs that would allow them to maintain copies of
the ledger directly, such that they would not need to query any particular MSB
with specific requests for information. Although the ledger entries themselves
would generally not contain metadata concerning the counterparties, the MSB
that submitted each transaction would be known to authorities, and it is
assumed that MSBs would maintain records of the transactions, including
transaction size and whatever information they have about the counterparties
even if it is limited, and that authorities would have access to such records.
Next, we specify who can access the ledger:

\begin{itemize}

\item \cz{Writing to the ledger.} We envision that the only entities
authorised to write to the ledger shall be the operators of the ledger, namely
the regulated MSBs (including but not limited to banks) and the central bank
itself. The central bank shall write the entries that create or destroy CBDC,
and MSBs shall write the entries that ``move'' tokens within the system by
signing them over from one keyholder to another. All entries would be approved
via a consensus mechanism in which all entries would be approved by a
supermajority (perhaps nearly all, depending upon the specific design) of the
private-sector participants.

\item \cz{Reading the ledger.} We envision that the set of entities authorised
to read the entries on the ledger shall include those who can write to the
ledger and, by extension, the regulators who oversee the parties that are
authorised to write to the ledger. We do not anticipate that a public-facing
API to read the ledger would be necessary, although a government might want to
provide such a mechanism, for example to streamline public oversight of the
system or to facilitate the investigation of suspicious activity. Privacy by
design would require that the transaction record on the ledger would not
generally provide transaction histories for individual users, so it is assumed
that there are limits to what the ledger entries would reveal.

\end{itemize}

\subsection{Non-Custodial Wallets}

Another important feature of our proposed architecture is \textit{privacy by
design}. Although we argue that data protection is no substitute for privacy
(see Section~\ref{ss:privacy}), Ulrich Bindseil notes that ``others will argue
that a more proportionate solution would consist in a sufficient protection of
electronic payments data''~\cite{bindseil2020}. In the case of our proposed
design, we might imagine that because the entire network is operated by
regulated MSBs, some researchers might suggest creating a ``master key'' or
other exceptional access mechanism~\cite{li2021} to allow an authority to break
the anonymity of retail CBDC users. The temptation to build exceptional access
mechanisms should be resisted, with appreciation for the history of such
arguments~\cite{abelson1997,abelson2015,benaloh2018} and subsequent
acknowledgement by policymakers in Europe and
America~\cite{hr5823,thomson2016}, who have repeatedly cited their potential
for abuse, as well as their intrinsic security vulnerabilities. Ultimately,
substituting data protection for privacy risks creating a dragnet for
law-abiding retail CBDC users conducting legitimate activities, and it will
never be possible for a data collector to prove that data have not been subject
to analysis. To force people to use a system that relies on data protection is
to attempt to manufacture trust, which is impossible; trust must be earned.
Furthermore, criminals and those with privilege will have a variety of options,
including but not limited to proxies, cryptocurrencies, and identity theft,
available to them as ``outside solutions'' in the event that lawmakers attempt
to force them into transparency.

Unlike designs that contain exceptional access mechanisms that allow
authorities to trace the counterparties to every transaction and, therefore, do
not achieve anonymity at all, our approach actually seeks to deliver true but
``partial'' anonymity, wherein the counterparties to a transaction can be
anonymous, but all transactions are subject to control at the interface with the
MSB. We believe that our design is unique in that it achieves both anonymity
and control by ensuring that all transactions involve a regulated actor but
without giving authorities (or insiders, attackers, and so on) the ability to
unmask the counterparties to transactions, either directly or via correlation
attacks.

To satisfy the requirement for privacy by design, we introduce the concept of a
\textit{non-custodial wallet}, which is software that interacts with the ledger
via an MSB that allows a retail CBDC user to unlink her CBDC tokens from any
meaningful information about her identity or the identity of any previous
owners of the tokens. A user would withdraw tokens from an MSB into her
non-custodial wallet and, after some length of time, return them to an MSB in a
subsequent transaction, as shown in Figure~\ref{f:withdrawal}. The ledger
system is operated as a \textit{public permissioned DLT system} in which
participants are regulated MSBs. Alice withdraws digital tokens from an MSB
into her non-custodial wallet in transaction $T_{out}$ and subsequently returns
them to an MSB in transaction $T_{in}$. The MSB from which the tokens are
withdrawn might or might not be the same as the MSB to which the tokens are
returned. Specifically, a transaction in which a fungible token flows from a
non-custodial wallet to an MSB reveals no meaningful information about the
history of the token or its owner.

\begin{figure}
\begin{center}
\begin{tikzpicture}[>=latex, node distance=3cm, font={\sf \small}, auto]\ts
\node (w1) at (0, 4) [] {
  \scalebox{0.08}{\includegraphics{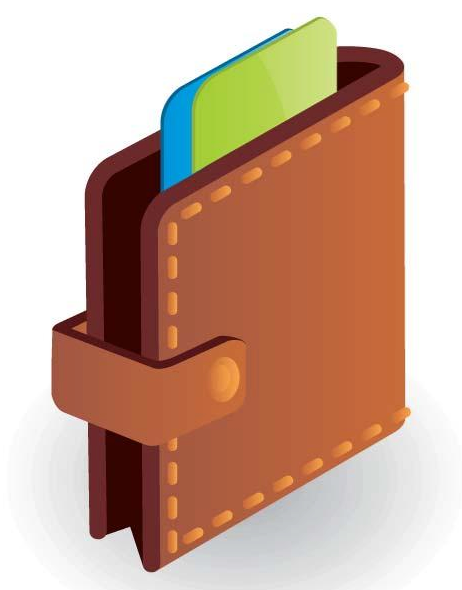}}
};
\node (a1) at (-3,2.5) [noshape, text width=4em] {
  \scalebox{0.2}{\includegraphics{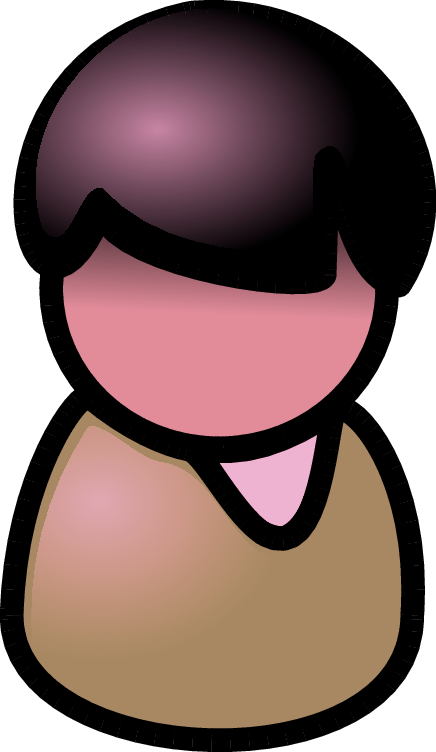}}
};
\node (a2) at (3,2.5) [noshape, text width=4em] {
  \scalebox{0.2}{\includegraphics{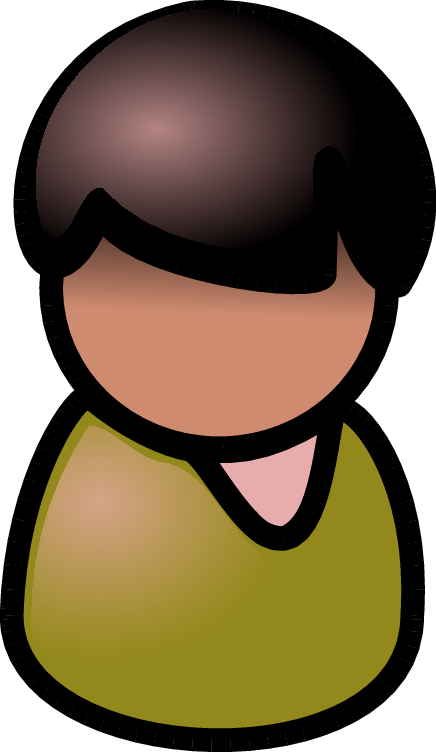}}
};
\node (b1) at (-2,1) [noshape, text width=4em] {
  \scalebox{0.4}{\includegraphics{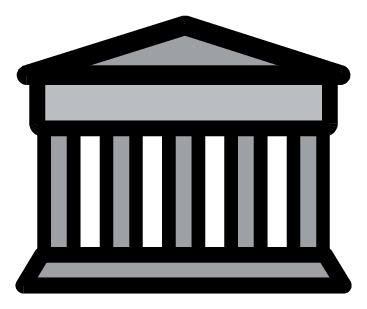}}
};
\node (b2) at (-2,-1) [noshape, text width=4em] {
  \scalebox{0.4}{\includegraphics{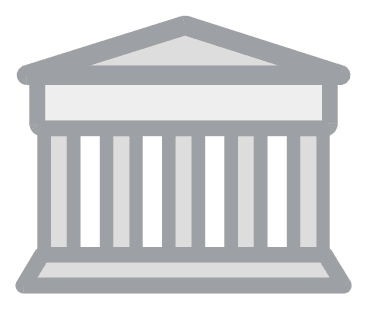}}
};
\node (b3) at (0,2.1) [noshape, text width=4em] {
  \scalebox{0.4}{\includegraphics{images/primary_bank_light.png}}
};
\node (b4) at (0,-2.1) [noshape, text width=4em] {
  \scalebox{0.4}{\includegraphics{images/primary_bank_light.png}}
};
\node (b5) at (2,1) [noshape, text width=4em] {
  \scalebox{0.4}{\includegraphics{images/primary_bank.png}}
};
\node (b6) at (2,-1) [noshape, text width=4em] {
  \scalebox{0.4}{\includegraphics{images/primary_bank_light.png}}
};
\node (c0) at (0,0) [
  circ,
  draw,
  line width=1.2mm,
  color=orange,
  minimum height=3.5cm,
  minimum width=3.5cm
] {};
\node (c1) at (0,0) [circ, draw, thick, minimum height=3.4cm, minimum width=3.4cm] {};
\node (c2) at (0,0) [circ, draw, thick, minimum height=3.6cm, minimum width=3.6cm] {};
\node (t1) at (0,0) [align=center] {Public\\\textbf{Permissioned}\\DLT System};
\node (t2) at (0,5) [align=center] {Alice's \textbf{Non-Custodial} Wallet*};
\node (t3) at (-3,3.5) [align=center] {Alice};
\node (t4) at (3,3.5) [align=center] {Bob};
\node (t5) at (-3.2,1) [align=center] {Alice's\\MSB};
\node (t6) at (3.2,1) [align=center] {Bob's\\MSB};
\draw[->, line width=0.9mm] (b1) edge[bend left=20] node {$T_{out}$} (w1);
\draw[->, line width=0.9mm] (w1) edge[bend left=20] node {$T_{in}$} (b5);
\end{tikzpicture}

\caption{\cz{Schematic representation of the system design.}}

\label{f:withdrawal}
\end{center}
\end{figure}

To support non-custodial wallets with the privacy features we describe, the
CBDC system must incorporate privacy-enhancing technology of the sort we
described in Section~\ref{ss:pets}. If we choose to use ZKP or a combination
of stealth addresses, Pedersen commitments, and ring signatures, then it might
be possible to avoid requiring receiving MSBs to immediately return the (spent)
tokens that they receive to the issuer. However, blind signatures might offer
some benefit in the form of computational efficiency, and the fact that we have
stipulated the involvement of an MSB in every transaction, they could be
appropriate for our use case. Specifically, because we stipulate that a token
withdrawn from a regulated MSB will be returned to a regulated MSB without
being transacted with other non-custodial wallets first, we can specify that a
user would first receive a blinded token from an MSB and later send the
unblinded version of the token to the recipient's MSB. Under the assumptions
made by Chaum, this can be done without revealing information that can be used
to link the transaction wherein the token is withdrawn to the transaction
wherein the token is returned.

It has been argued that modern cryptographic techniques, such as zero-knowledge
proofs, are too difficult to be understood or implemented effectively as part of
public infrastructure, although this view ignores the reality that such
cryptographic techniques are well-established. Additionally, there are many
instances of regulation that does not specify the details of the specific
technologies that are used to achieve compliance. Consider as an example the
co-regulatory approach taken by regulators in the context of best execution
networks, as described in Section~\ref{ss:governance}.

\subsection{User Engagement Lifecycle}

Figure~\ref{f:pevept} depicts a typical user engagement lifecycle with CBDC,
which we anticipate would be a typical use case for our design. This user
(Individual B) has a bank account and receives an ordinary payment via bank
transfer into her account (with Bank B). Then, the user asks her bank (Bank B)
to withdraw CBDC, which takes the form of a set of tokens that are transferred
to her non-custodial wallet via unlinkable transactions. On-ledger
transactions of CBDC are represented in the figure by the Pound Sterling symbol
(\pounds). (If Bank B had not received the CBDC directly from Bank A along
with the payment, then it might source the CBDC from its own holdings, or it
might receive the CBDC from the central bank in exchange for cash or reserves.)
Later, the user approaches a merchant or other service provider (Business C),
either in-person or online, that has an account with a bank (Bank C) that is
configured to receive CBDC. Using her non-custodial wallet, the user interacts
with software that facilitates an interaction between her non-custodial wallet
and the merchant's bank wherein the bank publishes a set of transactions to the
ledger effecting a transfer of CBDC from the user's non-custodial wallet to the
merchant's bank, credits the merchant's account, and informs the merchant that
the transaction was processed successfully. The merchant's bank then has the
option to return the CBDC to the central bank in exchange for cash or reserves.
The privacy features of the ledger design and the non-custodial wallet software
ensure that the user does not reveal anything about her identity or the history
of her tokens in the course of the transaction that can be used to identify her
or profile her behaviour. More generally, we envision that a retail user of
digital currency would receive it via one of four mechanisms:

\ifnum\x=0
  \end{paracol}
\fi
\begin{figure}
\widefigure
\begin{center}
\scalebox{1}{\begin{tikzpicture}[>=latex, node distance=3cm, font={\sf \small}, auto]\ts
\node (box1) at (8,0.2) [box, minimum width=2.8cm, minimum height=3.4cm] {};
\node (x1) at (-2,4) [align=center]{Human\\Layer};
\node (x2) at (-2,0) [align=center]{Token\\Layer};
\node (x3) at (-2,-4) [align=center]{Issuer\\Layer};
\node (r1) at (0,0) [noshape, text width=4em] {
  \scalebox{0.8}{\includegraphics{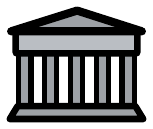}}
};
\node (s1) at (0.3,-0.2) [box, color=white, fill=magenta] {
  MSB
};
\node (c1) at (0,-4) [noshape, text width=4em] {
  \scalebox{0.8}{\includegraphics{images/primary_bank.pdf}}
};
\node (o1) at (0,4) [noshape, text width=4em] {
  \scalebox{1.0}{\includegraphics{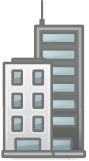}}
};
\node (r2) at (4,0) [noshape, text width=4em] {
  \scalebox{0.8}{\includegraphics{images/primary_bank.pdf}}
};
\node (s2) at (4.3,-0.2) [box, color=white, fill=magenta] {
  MSB
};
\node (c2) at (4,-4) [noshape, text width=4em] {
  \scalebox{0.8}{\includegraphics{images/primary_bank.pdf}}
};
\node (o2) at (4,4) [noshape, text width=4em] {
  \scalebox{0.5}{\includegraphics{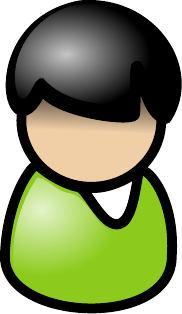}}
};
\node (r3) at (8,0) [noshape, text width=4em] {
  \scalebox{0.06}{\includegraphics{images/wallet-vector-xp.png}}
};
\node (o3) at (8,4) [noshape, text width=4em] {
  \scalebox{0.5}{\includegraphics{images/people-juliane-krug-08a.pdf}}
};
\node (r4) at (12,0) [noshape, text width=4em] {
  \scalebox{0.8}{\includegraphics{images/primary_bank.pdf}}
};
\node (s4) at (12.3,-0.2) [box, color=white, fill=magenta] {
  MSB
};
\node (c4) at (12,-4) [noshape, text width=4em] {
  \scalebox{0.8}{\includegraphics{images/primary_bank.pdf}}
};
\node (o4) at (12,4) [noshape, text width=4em] {
  \scalebox{1.0}{\includegraphics{images/office-towers.pdf}}
};
\node (m1) at (0.8,0.8) [noshape] {
  \scalebox{0.06}{\includegraphics{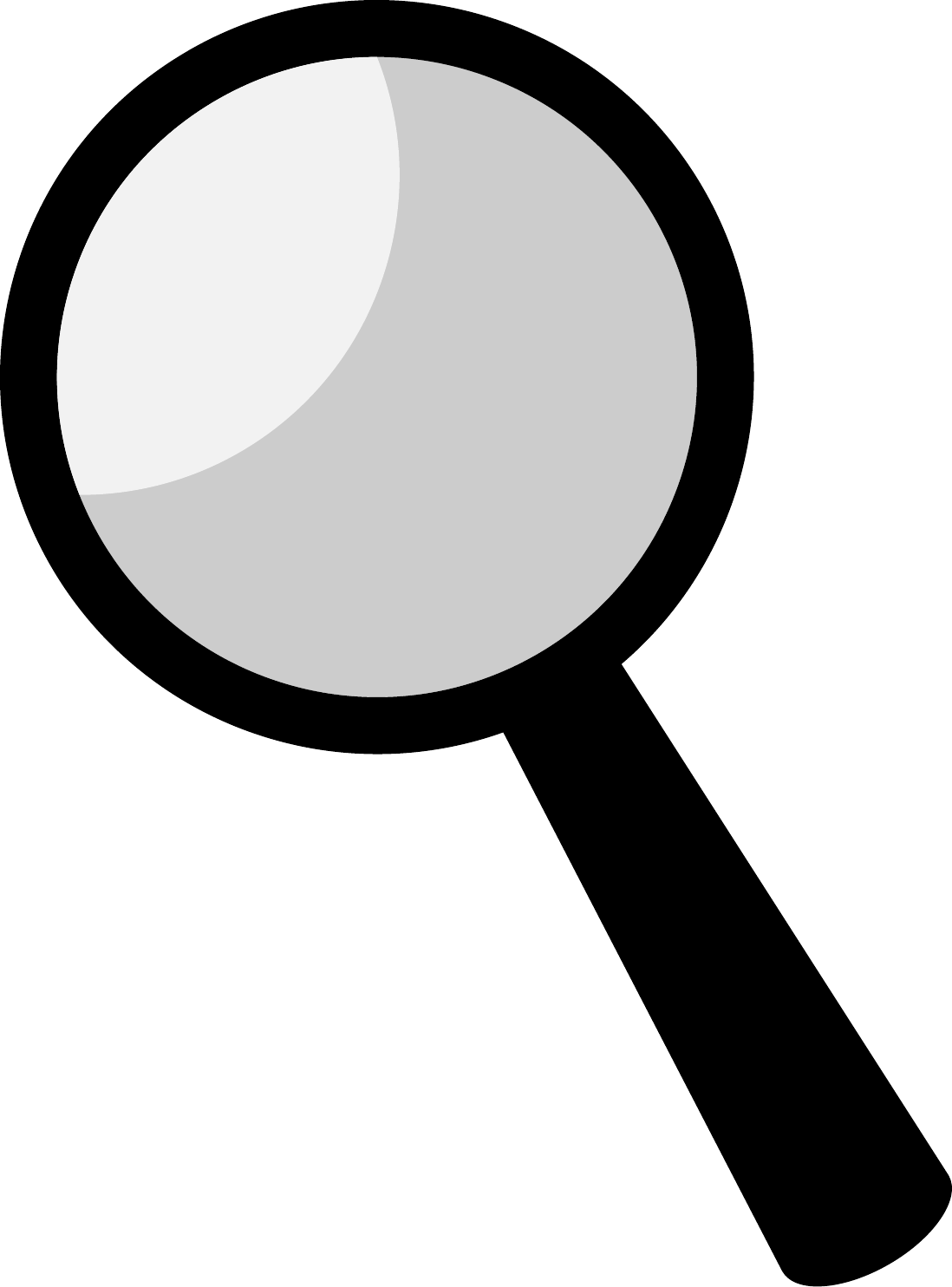}}
};
\node (m2) at (4.8,0.8) [noshape] {
  \scalebox{0.06}{\includegraphics{images/office-glass-magnify.pdf}}
};
\node (m2) at (12.8,0.8) [noshape] {
  \scalebox{0.06}{\includegraphics{images/office-glass-magnify.pdf}}
};
\node (m1) at (8.8,0.8) [noshape] {
  \scalebox{0.06}{\includegraphics{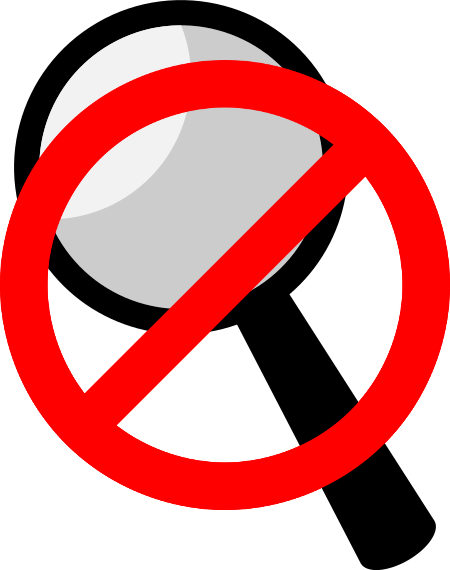}}
};

\draw[->, line width=0.5mm, align=center] (o1) -- node[below] {
  Payment\\(e.g. payroll)
} (o2);
\draw[->, line width=0.5mm, align=center] (o3) -- node[below] {
  Payment\\(e.g. purchase)
} (o4);
\draw[->, dashed, line width=0.5mm] (r1) -- node[below] {
  \textbf{\textit{\Huge \pounds}}
} (r2);
\draw[->, dashed, line width=0.5mm] (0, -3.4) -- node[right] {
  \textbf{\textit{\Huge \pounds}}
} (0, -1.2);
\draw[->, line width=0.5mm] (r2) -- node[below] {
  \textbf{\textit{\Huge \pounds}}
} (r3);
\draw[->, dashed, line width=0.5mm] (4, -3.4) -- node[right] {
  \textbf{\textit{\Huge \pounds}}
} (4, -1.2);
\draw[->, line width=0.5mm] (r3) -- node[below] {
  \textbf{\textit{\Huge \pounds}}
}(r4);
\draw[->, dashed, line width=0.5mm] (12, -1.2) -- node[right] {
  \textbf{\textit{\Huge \pounds}}
} (12, -3.4);
\node (d1) at (0,3) [nos] {Business A};
\node (d2) at (4,3) [nos] {Individual B};
\node (d3) at (8,3) [nos] {Individual B};
\node (d4) at (12,3) [nos] {Business C};
\draw[<->, line width=0.5mm, align=center] (0, 2.8) -- node[sloped, above] {
  account
} (0, 0.6);
\draw[<->, line width=0.5mm, align=center] (4, 2.8) -- node[sloped, above] {
  account
} (4, 0.6);
\draw[<->, line width=0.5mm, align=center] (8, 2.8) -- (8, 0.6);
\draw[<->, line width=0.5mm, align=center] (12, 2.8) -- node[sloped, above] {
  account
} (12, 0.6);
\node (e1) at (0,-0.8) [nos] {Bank A};
\node (e2) at (4,-0.8) [nos] {Bank B};
\node (e3) at (8,-1.0) [nos] {non-custodial wallet};
\node (e4) at (12,-0.8) [nos] {Bank C};
\node (f1) at (0,-4.8) [nos] {central bank};
\node (f2) at (4,-4.8) [nos] {central bank};
\node (f4) at (12,-4.8) [nos] {central bank};
\end{tikzpicture}}

\caption{\cz{Schematic representation of a typical user engagement
lifecycle.}}

\label{f:pevept}
\end{center}
\end{figure}
\ifnum\x=0
  \begin{paracol}{2}
  \linenumbers
  \switchcolumn
\fi

\begin{enumerate}

\item \cz{Via an exchange of money from an account with an MSB into digital
currency.} We stipulate that an individual or business with an account with an
MSB could opt to \textit{withdraw} digital currency from the account into a
non-custodial wallet. Digital currency held by a retail user in the user's
non-custodial wallet would be like cash. Because it is not held by an MSB, it
would not be invested and it would not earn true interest; it would be
tantamount to holding cash in a physical wallet. It might be possible for
governments to incentivise or penalise the asset itself, but this would not be
``true'' interest and would not serve the same purpose. Similarly, an
individual or business with an account with an MSB could opt to
\textit{deposit} digital currency from a non-custodial wallet into an account,
reversing the process, as shown in Figure~\ref{f:pevdeposit}. Retail users
would be permitted to deposit funds into their own accounts, possibly subject
to certain limits or additional checks in the event that such deposits are
frequent or large.

\begin{figure}
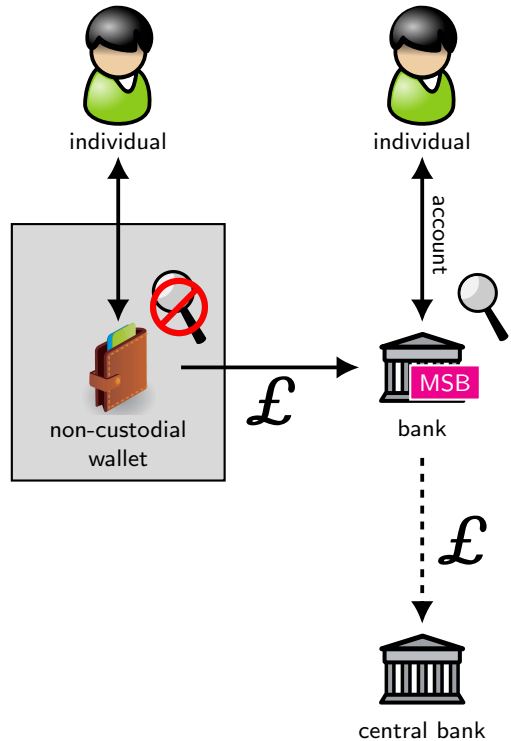

\begin{center}
\begin{tikzpicture}[>=latex, node distance=3cm, font={\sf \small}, auto]\ts
\node (box1) at (8,0.2) [box, minimum width=2.8cm, minimum height=3.4cm] {};
\node (r3) at (8,0) [noshape, text width=4em] {
  \scalebox{0.06}{\includegraphics{images/wallet-vector-xp.png}}
};
\node (o3) at (8,4) [noshape, text width=4em] {
  \scalebox{0.5}{\includegraphics{images/people-juliane-krug-08a.pdf}}
};
\node (r4) at (12,0) [noshape, text width=4em] {
  \scalebox{0.8}{\includegraphics{images/primary_bank.pdf}}
};
\node (s4) at (12.3,-0.2) [box, color=white, fill=magenta] {
  MSB
};
\node (c4) at (12,-4) [noshape, text width=4em] {
  \scalebox{0.8}{\includegraphics{images/primary_bank.pdf}}
};
\node (o4) at (12,4) [noshape, text width=4em] {
  \scalebox{0.5}{\includegraphics{images/people-juliane-krug-08a.pdf}}
};
\node (m2) at (12.8,0.8) [noshape] {
  \scalebox{0.06}{\includegraphics{images/office-glass-magnify.pdf}}
};
\node (m1) at (8.8,0.8) [noshape] {
  \scalebox{0.06}{\includegraphics{images/office-glass-magnify-no.png}}
};
\draw[->, line width=0.5mm] (r3) -- node[below] {
  \textbf{\textit{\Huge \pounds}}
}(r4);
\draw[->, dashed, line width=0.5mm] (12, -1.2) -- node[right] {
  \textbf{\textit{\Huge \pounds}}
} (12, -3.4);
\node (d3) at (8,3) [nos] {individual};
\node (d4) at (12,3) [nos] {individual};
\draw[<->, line width=0.5mm, align=center] (8, 2.8) -- (8, 0.6);
\draw[<->, line width=0.5mm, align=center] (12, 2.8) -- node[sloped, above] {
  account
} (12, 0.6);
\node (e3) at (8,-1.0) [nos] {non-custodial wallet};
\node (e4) at (12,-0.8) [nos] {bank};
\node (f4) at (12,-4.8) [nos] {central bank};

\end{tikzpicture}

\caption{\cz{Schematic representation of a user depositing CBDC into a bank
account.}}

\label{f:pevdeposit}
\end{center}
\end{figure}

\item \cz{As a recipient of digital currency from an external source,
received into an account with an MSB.} In this case, the user would be the
recipient of a digital currency payment. The sender of the payment might be
known, for example if it is an account with an MSB, or it might be unknown,
specifically if it is a non-custodial wallet.

\item \cz{As a recipient of digital currency from an external source, received
into a non-custodial wallet.} Any transaction in which a non-custodial wallet
receives digital currency from an external source must be mediated by an MSB,
so the key difference between this mode of receiving digital currency and a
withdrawal from the user's own account is that in this case the recipient does
not have (or is not using) an account with the MSB. This form of transaction
is illustrated in Figure~\ref{f:pevmed}. Retail CBDC users wishing to transact
with each other via their non-custodial wallets must transact via a regulated
institution or a regulated business with an account with a regulated
institution. The institution creates on-ledger transactions from the
non-custodial wallet of one retail CBDC user and to the non-custodial wallet of
another retail CBDC user without creating accounts for the retail CBDC users.
We imagine that there would be certain legal requirements, such as transaction
limits or a requirement for the recipient to provide positive identification
documents to a human clerk, that would govern the role of the MSB in such
transactions. We also imagine that this process could be particularly useful
as a means to deliver government payments (for economic stimulus or for other
reasons) to retail users without bank accounts, as illustrated in
Figure~\ref{f:pevstimulus}. This example shows how a retail user might claim
CBDC that she is eligible to receive, either directly from the central bank or
from an institution such as the State treasury or a private-sector bank. The
user would identify herself to a regulated MSB, which would carry out the
requisite compliance checks.

\begin{figure}
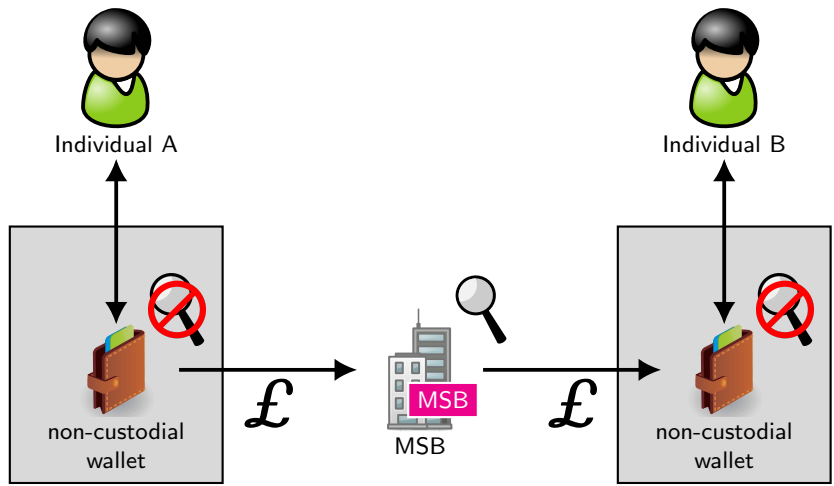

\begin{center}
\begin{tikzpicture}[>=latex, node distance=3cm, font={\sf \small}, auto]\ts
\node (box1) at (0,0.2) [box, minimum width=2.8cm, minimum height=3.4cm] {};
\node (box1) at (8,0.2) [box, minimum width=2.8cm, minimum height=3.4cm] {};
\node (r1) at (0,0) [noshape, text width=4em] {
  \scalebox{0.06}{\includegraphics{images/wallet-vector-xp.png}}
};
\node (o1) at (0,4) [noshape, text width=4em] {
  \scalebox{0.5}{\includegraphics{images/people-juliane-krug-08a.pdf}}
};
\node (r2) at (4,0) [noshape, text width=4em] {
  \scalebox{1.0}{\includegraphics{images/office-towers.pdf}}
};
\node (s2) at (4.3,-0.4) [box, color=white, fill=magenta] {
  MSB
};
\node (r3) at (8,0) [noshape, text width=4em] {
  \scalebox{0.06}{\includegraphics{images/wallet-vector-xp.png}}
};
\node (o3) at (8,4) [noshape, text width=4em] {
  \scalebox{0.5}{\includegraphics{images/people-juliane-krug-08a.pdf}}
};
\node (m1) at (0.8,0.8) [noshape] {
  \scalebox{0.06}{\includegraphics{images/office-glass-magnify-no.png}}
};
\node (m2) at (4.8,0.8) [noshape] {
  \scalebox{0.06}{\includegraphics{images/office-glass-magnify.pdf}}
};
\node (m3) at (8.8,0.8) [noshape] {
  \scalebox{0.06}{\includegraphics{images/office-glass-magnify-no.png}}
};

\draw[->, line width=0.5mm] (r1) -- node[below] {
  \textbf{\textit{\Huge \pounds}}
}(r2);
\draw[->, line width=0.5mm] (r2) -- node[below] {
  \textbf{\textit{\Huge \pounds}}
}(r3);
\node (d1) at (0,3) [nos] {Individual A};
\node (d3) at (8,3) [nos] {Individual B};
\draw[<->, line width=0.5mm, align=center] (0, 2.8) -- (0, 0.6);
\draw[<->, line width=0.5mm, align=center] (8, 2.8) -- (8, 0.6);
\node (e1) at (0,-1.0) [nos] {non-custodial wallet};
\node (e2) at (4,-1) [nos] {MSB};
\node (e3) at (8,-1.0) [nos] {non-custodial wallet};

\end{tikzpicture}

\caption{\cz{Schematic representation of a mediated transaction between
consumers.}}

\label{f:pevmed}
\end{center}
\end{figure}

\begin{figure}
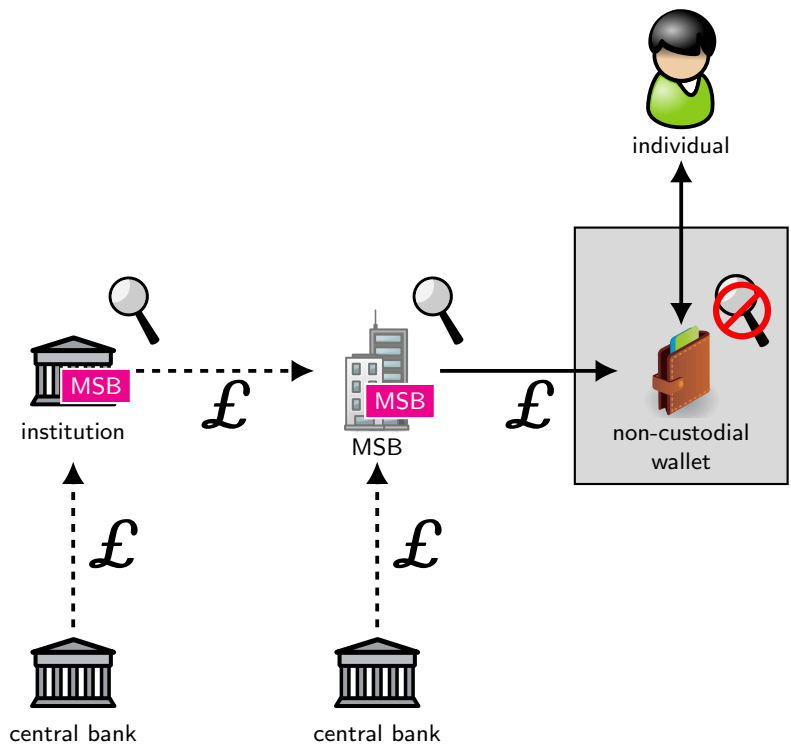

\begin{center}
\begin{tikzpicture}[>=latex, node distance=3cm, font={\sf \small}, auto]\ts
\node (box1) at (8,0.2) [box, minimum width=2.8cm, minimum height=3.4cm] {};
\node (r2) at (4,0) [noshape, text width=4em] {
  \scalebox{1.0}{\includegraphics{images/office-towers.pdf}}
};
\node (s2) at (4.3,-0.4) [box, color=white, fill=magenta] {
  MSB
};
\node (c0) at (0,-4) [noshape, text width=4em] {
  \scalebox{0.8}{\includegraphics{images/primary_bank.pdf}}
};
\node (c1) at (0,0) [noshape, text width=4em] {
  \scalebox{0.8}{\includegraphics{images/primary_bank.pdf}}
};
\node (r3) at (8,0) [noshape, text width=4em] {
  \scalebox{0.06}{\includegraphics{images/wallet-vector-xp.png}}
};
\node (c2) at (4,-4) [noshape, text width=4em] {
  \scalebox{0.8}{\includegraphics{images/primary_bank.pdf}}
};
\node (o3) at (8,4) [noshape, text width=4em] {
  \scalebox{0.5}{\includegraphics{images/people-juliane-krug-08a.pdf}}
};
\node (m1) at (0.8,0.8) [noshape] {
  \scalebox{0.06}{\includegraphics{images/office-glass-magnify.pdf}}
};
\node (m2) at (4.8,0.8) [noshape] {
  \scalebox{0.06}{\includegraphics{images/office-glass-magnify.pdf}}
};
\node (m3) at (8.8,0.8) [noshape] {
  \scalebox{0.06}{\includegraphics{images/office-glass-magnify-no.png}}
};

\draw[->, dashed, line width=0.5mm] (c1) -- node[below] {
  \textbf{\textit{\Huge \pounds}}
}(r2);
\draw[->, line width=0.5mm] (r2) -- node[below] {
  \textbf{\textit{\Huge \pounds}}
}(r3);
\node (d3) at (8,3) [nos] {individual};
\draw[<->, line width=0.5mm, align=center] (8, 2.8) -- (8, 0.6);
\node (e1) at (0,-0.8) [nos] {institution};
\node (s1) at (0.3,-0.2) [box, color=white, fill=magenta] {
  MSB
};
\node (e2) at (4,-1) [nos] {MSB};
\node (e3) at (8,-1.0) [nos] {non-custodial wallet};
\draw[->, dashed, line width=0.5mm] (0, -3.4) -- node[right] {
  \textbf{\textit{\Huge \pounds}}
} (0, -1.2);
\draw[->, dashed, line width=0.5mm] (4, -3.4) -- node[right] {
  \textbf{\textit{\Huge \pounds}}
} (4, -1.2);
\node (f1) at (0,-4.8) [nos] {central bank};
\node (f2) at (4,-4.8) [nos] {central bank};

\end{tikzpicture}

\caption{\cz{Schematic representation of a disbursement to a retail user with a
non-custodial wallet.}}

\label{f:pevstimulus}
\end{center}
\end{figure}

\item \cz{Via an exchange of physical cash into digital currency.} The
transaction in which physical cash is converted to digital currency would be
facilitated by an MSB, subject to appropriate rules, just as in the case that
digital currency is received directly from an external source. For example,
the MSB might be required to ask for information concerning the origin of the
cash if the amount exceeds a certain threshold.

\end{enumerate}

Note that retail bank accounts are not generally expected to hold CBDC on
behalf of a particular user, any more than retail bank accounts would hold cash
on behalf of a particular user. A bank would swap CBDC for central bank
reserves from time to time, and vice-versa, with the expectation that the bank
would furnish CBDC to its retail customers, subject to limits on the size and
rate of withdrawals.

Note also that the messages on the ledger are published by regulated financial
institutions. This is an important feature of the system design: all
transactions on the ledger must be published by a regulated MSB, and because
the ledger is operated entirely by regulated MSBs, private actors cannot
exchange value directly between their non-custodial wallets. The non-custodial
wallets offer a layer of indirection wherein MSBs would not be able to identify
the \textit{counterparties} to the transactions involving non-custodial
wallets. Every non-custodial wallet transaction is unlinkable to every other
non-custodial wallet transaction, up to the limit of circumstantial evidence,
such as timing. Banks might be required to know their customers, but merchants
generally do not. Furthermore, a merchant's bank does not need to know the
merchant's customers, and a merchant's customer's bank does not need to know
about the merchant or its bank at all. For instances wherein merchants really
do need to know their customers, the reason is generally about the substance of
the relationship rather than the mechanism of the payment, and identification
of this sort should be handled outside the payment system.

By providing a mechanism by which no single organisation or group would be able
to build a profile of any individual's transactions in the system, the use of a
distributed ledger achieves an essential requirement of the design. In
addition to our previously stated requirement that transactions into and out of
the non-custodial wallets would be protected by mechanisms, such as stealth addresses
or zero-knowledge proofs, to disentangle the outflows from the inflows,
individuals would be expected to use their non-custodial wallets to transact with
many different counterparties, interacting with the MSBs chosen by their
counterparties and not with the MSBs from which their non-custodial wallets were
initially funded.

Figure~\ref{f:pevmed} depicts the mechanism by which individuals would transact
from one non-custodial wallet to another. They must first identify a regulated MSB
to process the transaction onto the ledger, perhaps in exchange for a small
fee. The MSB would process a set of transactions from the first non-custodial wallet
to the MSB and from the MSB to the second non-custodial wallet. An MSB could provide
a similar service for an individual exchanging CBDC for cash, or vice-versa.
Presumably, the MSB would gather whatever information is needed from its
customers to satisfy compliance requirements, although we imagine that strong
client identification, such as what might conform to the FATF
recommendations~\cite{fatf-recommendations}, could be waived for transactions
that take place in-person and are sufficiently small. In the case of small
online transactions between two persons, we imagine that an attribute-backed
credential indicating that either the sender or the receiver is eligible to
transact might be sufficient~\cite{goodell2019a}. Finally, some MSBs could
provide token-mixing services for retail CBDC users who had accidentally exposed
metadata about the tokens in their non-custodial wallets.

Concerning the hypothetical stimulus described in Figure~\ref{f:pevstimulus},
we note that if a government intends to make stimulus payments to a specific
set of eligible individuals,\footnote{The government could do the same for
businesses, if desired.} notwithstanding the possibility that this set might
include all citizens or residents, then it could refer to each such individual
using a unique taxpayer identification number. Then, the government could ask
each eligible party to specify a bank account, current account, or wallet into
which to deposit the funds. This approach might work in many cases, although
it might not work for eligible individuals or businesses without bank accounts.
To address the gap, the government could ask eligible parties to identify
themselves to a qualified MSB for verification, for example a post office, that
would be able to carry out the required identification procedures to determine
whether the prospective recipient has the right to make a claim associated with
a particular taxpayer identification number. Once this is done, the MSB could
enter a transaction that delivers the digital currency to the individual's
non-custodial wallet directly, avoiding the need for a bank account. We propose that
each of these options could be provided to both individuals and businesses.

\subsection{Security Considerations}

Since digital currencies generally rely upon the use and management of
sensitive cryptographic information, such as keys, we recognise that a digital
currency that allows users to hold tokens outside of the protection of an
account with a financial institution would also introduce responsibility on the
part of users to manage the security of those tokens. Users have a range of
possible options at their disposal, including encrypted devices with one-factor
or two-factor authentication, third-party custodial services, single-use
physical tokens as an alternative to wallet software for their general-purpose
devices, and simply choosing to limit the amount of digital currency that they
hold at any moment. We suggest that all of these approaches could be useful,
and as with many financial decisions, the best choice would be a function of
the preferences and risk profile of each individual user.

Of course, privacy-enhancing technology alone does not imply perfect anonymity.
Although the approach we describe intentionally avoids linking transactions to
each other or to any identifiers that might be associated with the owners of
non-custodial wallets, circumstantial evidence might potentially link multiple
transactions to each other, and information that links a user to a particular
transaction might be combined with other data to deduce other patterns of use.
For example, users seeking to make anonymous transactions via the Internet
might want to avoid revealing information, such as network carrier information,
that could be used to de-anonymise them. In addition, timing attacks are a
risk characteristic of any low-latency activity, and we would strongly
recommend that users wait for a period of time, perhaps hours or days, between
withdrawing funds into their non-custodial wallets and spending those funds.
If most users were to adopt the practice of withdrawing CBDC into their
non-custodial wallets and then spending it immediately, then the benefit of
blending in with other users whose money is waiting within non-custodial
wallets would be lost.

We imagine that under some circumstances an individual might decide to share
the private cryptographic information (e.g., a private key that can be used to
initiate a transaction) associated with digital currency with another
individual, thereby allowing the other individual to transact it on her behalf.
This kind of sharing of privileges might be appropriate in the same way that
colleagues or family members might share debit cards. We do not consider that
such an exchange of information would constitute a payment, since there is
nothing intrinsic to the system that would stop the first party from spending
the digital currency before the second party has a chance to do so. It would
be appropriate to characterise such an exchange as a ``shared wallet'' or a
``promise of payment'' rather than a payment itself, similar to providing a
post-dated cheque, and there is no mechanism to prevent people from making
promises to each other. Once an individual or business is in possession of
digital currency, the ways to dispose of the digital currency are the inverses
of the methods to acquire it.

\section{Analysis}
\label{s:analysis}

We note that although it can accommodate CBDC, the digital currency system we
propose can be generalised as a ``value container''~\cite{pilkington2016} that
can be extended to potentially represent a plethora of different assets and
their underlying infrastructure, including but not limited to central bank or
government assets. For the purpose of our analysis, we focus on the use of our
proposed design for CBDC and, specifically, retail CBDC, as a means of allowing
the general public to have broad access to an public, digital form of cash.

\subsection{CBDC as a Retail Payment System}
\label{ss:retail}

We suggest that a primary benefit of CBDC is its ability to be held in
non-custodial wallets by retail users. The argument that CBDC should be held
only in custodial accounts actually follows from two assumptions: first, that it
is not possible to remunerate tokenised assets directly; and second, that the
purpose of CBDC is primarily to solve a problem of efficiency, for example of
transaction costs or monetary policy transmission, and nothing more. However,
there are plausible mechanisms that can remunerate tokenised assets directly,
and the inexorable decline in cash as a means of payment presents a problem
that is manifestly deeper than monetary policy transmission. Thanks to cash,
people have always had the ability to conduct financial transactions using
assets that they could control completely, for which their spending habits
cannot be profiled, and which are not generally subject to discrimination or
interception by third parties. However, the decline in cash use suggests that
cash infrastructue might soon become economically untenable, in which case
these foundational rights face elimination by default. Therefore, CBDC can be
seen, perhaps first and foremost, as an opportunity to allow retail users to
continue to enjoy the benefits of accountless money in the digital age.

We ask whether CBDC is best seen as a modern form of bank deposits or as a
digital form of cash. If CBDC were to be account-based and suitable for
rehypothecation, then it might plausibly substitute for bank deposits in the
general case, although if, as we propose, CBDC were to be token-based and not
suitable for rehypothecation, then it would be much more cash-like. In the
latter case, users would still have reasons, including interest and inflation
risk, to continue to prefer bank deposits as a store of value and to use CBDC
principally as a means of payment, even if both forms of money were usable for
both purposes.

Importantly, the architectural features of our proposal make it private
\textit{by design and by default}. Our proposal shows how a measure of true
anonymity can be maintained even with an institutionally operated platform,
thus disrupting the notion that electronic, institutionally supported retail
payment methods must necessarily capture all available data about transacting
parties.

\subsection{Decentralisation}

There are some important questions to ask about a token-based design, including
whether we need the tokens to be issued by the central bank directly, or by
other institutions (``stablecoins''), or whether the tokens can operate
entirely outside the institutional milieu (``cryptocurrency''). We note that
stablecoins introduce systemic risk. Their design relies upon a peg to some
other asset, which can ultimately be undone. Users of the stablecoin,
therefore, incur counterparty risk to those who are tasked with maintaining the
peg. This counterparty risk implies either that the stablecoin must trade at a
discount to the asset to which it is pegged, or that the peg would be
underwritten by a government actor, such as a central bank. In the former case,
the stablecoin is not so stable. In the latter case, the stablecoin is not
really different from fiat currency.

Token-based systems, including systems with strong privacy characteristics, can
be centralised, relying upon a specific arbiter to handle disputes about the
validity of each transaction (possibly with a different arbiter for different
transactions), or they can be decentralised, using a distributed ledger to
validate each transaction \textit{ex ante} via a consensus process. For a
decentralised design, we consider the question of who the system operators
would be. In the case of CBDC, for example, although we assume that the
central bank would be responsible for the design and issuance of CBDC tokens,
we do not make the same assumption about the responsibility for the operation
of a transaction infrastructure or payment system, which historically has
generally been operated by private-sector organisations. As mentioned earlier,
systems for payments, clearing, and settlement are often a collaborative
effort~\cite{bis2012,bis2012a}. Indeed, modern digital payments infrastructure
based on bank deposits depends upon a variety of actors, and we imagine that
digital payments infrastructure based on CBDC would do so, as well. The
responsibility to manage and safeguard the value of currency is not the same as
the responsibility to manage and oversee transactions, and the responsibility
to supervise payment systems is not the same as the responsibility to operate
them. A design that externalises responsibility for the operation of a
transaction infrastructure supporting CBDC is not incompatible with the
operational role of a central bank in using CBDC to create money and implement
monetary policy.

The CBDC proposed in our design model relies upon the DLT infrastructure for a
variety of reasons. In our view, this is currently the most plausible method
of implementation whereby the central bank can collaborate with private sector
firms, via either public-private partnerships or other collaborative and
supervisory models, to deliver a national payments infrastructure operated by
the private sector. The use of DLT does not imply that households and retail
members of the public must have a direct account or relationship with the
central bank, as some authors have wrongly assumed. On the contrary, our
design recognises the important role of MSBs, especially for identifying,
onboarding, and registering new customers, satisfying compliance requirements,
and managing their accounts, if applicable.

In our view, the benefits of DLT broadly fall into three categories, all of
which relate to the scope for errors, system compromise, and potential
liability arising from exogenous or endogenous risk scenarios. We believe that
each of these benefits is indispensable and that all of them are necessary for
the system to succeed:

\begin{enumerate}

\item\cz{Eliminating the direct costs and risks associated with operating a
live system with a role as master or the capacity to arbitrate.} Because its
database is centrally managed, a centralised ledger would necessarily rely upon
some central operator that would have an operational role in the transactions.
This operational role would have the following three implications. First, the
central operator would carry administrative responsibility, including the
responsibility to guarantee system reliability on a technical level and handle
any exceptions and disputes on both a technical and human level. Second,
because the central operator would be positioned to influence transactions, it
would incur the cost of ensuring that transactions are carried out as expected,
as well as the risk of being accused of negligence or malice, whether or not
they are carried out as expected. Third, because the central operator
unilaterally determines what is allowed and what is not, it might be accused of
failing to follow the established rules.

\item\cz{Preventing unilateral action on the part of a single actor or group.}
Following the argument of Michael Siliski~\cite{siliski2018}, the administrator
of a centralised ledger could ban certain users or favour some users over
others; implicitly or explicitly charge a toll to those who use the system;
tamper with the official record of transactions; change the rules at any time;
or cause it to stop functioning without warning.

\item\cz{Creating process transparency and accountability for system
operators.} Because the administrator of a centralised ledger can make
unilateral decisions, there is no way for outside observers to know whether it
has carried out its responsibilities directly. In particular, its management
of the ledger and the means by which other parties access the ledger are under
its exclusive control, and the administrator has no need to publicise its
interest in changing the protocol or ask others to accept its proposed changes.
With DLT, it is possible to implement \textit{sousveillance} by ensuring that
any changes to the rules are explicitly shared with private-sector operators.

\item\cz{Improving efficiency and service delivery through competition and
scope for innovation.} Vesting accountability for system operation in
operators who are incentivised to perform would make it possible to achieve
important service delivery objectives, ranging from adoption in the first
instance to financial inclusion and non-discrimination, through private-sector
incentives (e.g., supporting local banks) rather than top-down political
directives.

\end{enumerate}

Each of these advantages of DLT relates to the scope for errors, system
compromise, and potential liability arising from exogenous or endogenous risk
factors surrounding a central authority. Although a central regulator might
have oversight over the operation of the transaction network and might decide
which private-sector MSBs are eligible to participate, DLT makes it possible to
assign responsibility for the transactions to the MSBs themselves.
Specifically, an MSB is responsible for each transaction that it writes to the
ledger, and the DLT can be used to create a (potentially) immutable record
binding each transaction to the corresponding MSB that submitted it, without
the need for a central actor would to be responsible for individual
transactions.

\subsection{Impact on Liquidity}
\label{ss:liquidity}

The issuance and use of CBDC could become a useful tool for central banks in
managing aggregate liquidity. For example, were CBDC to be widely held and
adopted for use, it could lead to a shift in aggregate liquidity, which refers
to the assets being used and exchanged and which carry a liquidity
premium~\cite{10}. Under certain models, a CBDC would lead to efficient
exchange, particularly given that it is a low cost medium of exchange and has a
stable unit of account, and particularly in the case wherein the digital
currency (as we propose it) is being used in a broad range of decentralised
transactions, and allows for monetary policy transmission channels on trading
activity to be strengthened. The central bank would have at its disposal
certain capabilities in controlling the supply and price of CBDC, including
through the use of (dis)incentives to generate a higher liquidity or lower
premium in CBDC and in bank deposits, subject to where investment frictions
exist in a much more targeted way~\cite{10}. Moreover, CBDC can be used as
intraday liquidity by its holders, whereas liquidity-absorbing instruments
cannot achieve the same effect. At present, there are few short-term money
market instruments that inherently combine the creditworthiness and the
liquidity that a CBDC could potentially provide. CBDC, therefore, could play
an important deterrent role against liquidity shocks.

One possible concern about CBDC is that individuals might run from bank
deposits to CBDC during a financial crisis. Although such a run is
conceivable, we argue that it is no more likely with our proposed system for
CBDC than it is with cash. Indeed, a CBDC could support replacing private
sector assets into risk-free assets to address the need for safe assets,
particularly given that although bank deposits are broadly insured up to some
amount, they continue to exhibit credit and residual liquidity risks. At the
same time, however, limits to the rate at which CBDC can be withdrawn from
financial institutions, and we imagine that individuals would be subject to
limits on their withdrawals of CBDC from their bank accounts, just as they are
subject to limits on their withdrawals of cash. If a run were underway, its
pace would be limited by such limits, and in principle, the government could
even ask banks to impose tighter limits or to disallow withdrawals from banks
entirely in the event of an emergency. Moreover, if the government chooses to
guarantee bank deposits up to an amount, then the other benefits afforded by
such deposits coupled with that guarantee would disincentivise such a run. In
other instances, the cost-benefit and risk-reward profile would require more
specific analysis on a jurisdiction by jurisdiction basis. Furthermore,
expiration dates on digital assets, and the lack of interest payments can
strongly disincentivise hoarding. Because we recognise significant utility for
bank deposits even in the presence of CBDC, we suggest that CBDC would be be
complementary to deposits and that banks would play a fundamental role in the
issuance and storage of CBDC tokens.

\subsection{Impact on the Financial Industry}

Our proposal frames CBDC as a distinct financial instrument but one that
nonetheless shares many features with cash, including being fully
collateralised and not providing for the ability to lend or rehypothecate. In
essence, it would strictly remain M0 money. Moreover, we are not proposing a
subordinate role for banknotes, nor for bank deposits. On the contrary, we
understand all three instruments to have merit and value to households and
firms within an economy and can be used to complement one another and increase
the overall welfare of individuals and firms through the adoption of
CBDC~\cite{21}. An example of the inherent difficulties within proposals that
argue for the abolition of cash is that the increase in its use is
predominantly situated within lower socioeconomic segments of a community, and
using CBDC to drive out cash would adversely impact those households and firms.
One question to ask is whether the cost of maintaining cash infrastructure
outweighs the cost of providing universal access to technology infrastructure
that would replace it.

The most direct impact of our approach to digital currency on the financial
industry involves risk management, on several levels. By improving the speed
of settlement, digital currency can be used to facilitate liquidity risk
management among financial institutions. Digital currency can also be used to
address systemic risk, both explicitly, by offering regulators a view into
substantially every transaction, as well as implicitly, by offering governments
a tool to implement stimulus while controlling the aggregate leverage in the
system.

Considering that, in general, DLT offers a promising risk-mitigation tool
\cite{TascaMorini}, our design relies on a DLT network operated by MSBs and
other private-sector institutions rather than a centralised ledger run by a
single public (or private, like in all the stablecoin solutions)
organisation. As such, our approach addresses a variety of risks associated
with relying upon a central arbiter: (1) technical risks associated with
availability, reliability, and maintenance; (2) risks associated with trust and
operational transparency; and (3) financial and legal risks. Our approach also
allows the private sector to operate the infrastructure for retail payments,
clearing, and settlement, while allowing government regulators to oversee the
system at an organisational level. Because we imagine that digital currency
will complement rather than substitute for bank deposits, our approach
leverages the role of commercial banks without forcibly decreasing their
balance sheets. In particular, because we believe that the main purpose of
CBDC tokens will be to facilitate electronic payments rather than to serve as a
long-term store of value, we do not anticipate that the balance sheets of
central banks will increase significantly as a result of its introduction.

\subsection{Impact on Fraud and Tax Evasion}
\label{ss:fraud}

We imagine that a rigorous compliance regime will govern the behaviour of MSBs
and the relationships they have with their customers. We assume that banks, in
particular, will have requirements for strong customer identification, and other
MSBs, such as wire transfer firms, currency exchanges, and post offices, will
face a combination of transaction limitations and procedures for identification
and authorisation. We assume that authorities will be able to see every
transaction that takes place, as well as the specific MSB that creates that
transaction, and we also assume that authorities will have access to the
records that the MSBs are required to maintain concerning the transactions they
facilitate.

Nevertheless, because our system allows a measure of true anonymity, it does
not provide a way to reveal the identities of both counterparties to
authorities. In particular, even if authorities have all of the records, some
transactions will have non-custodial wallets as a counterparty, just as some cash
transactions have anonymous counterparties. Although authorities might know
all of the retail users and their history of digital currency withdrawals, they
will not be able to link a non-custodial wallet to a specific retail user. Recall
that retail users will be able to withdraw digital currency from an MSB in the
same manner that they would withdraw cash from a bank or ATM, with similar
limits and restrictions. Retail users would be able to spend digital currency
the same way that they would be able to spend cash, making purchases with
vendors who are also subject to limits and restrictions, as well as profiling by
their financial institutions, and who know that their receipt of tokens will be
monitored by authorities. Authorities would know who had recently withdrawn
digital currency into a non-custodial wallet just as they would know who had recently
withdrawn cash, and they would also know who had recently received digital
currency from a non-custodial wallet. However, it would not be possible to use the
digital currency to link a specific recipient of cash to a specific
counterparty that had made a withdrawal. We argue that this property of cash
is necessary and fundamental to protect retail users from profiling and
manipulation by adversaries and other powerful interests including private
sector participants. Furthermore, revealing mutual counterparty information
for every transaction would divert the onus of fraud detection to law
enforcement agencies, effectively increasing their burden, while well-motivated
criminals would still be able to use proxies or compromised accounts to achieve
their objectives, even if every transaction were fully transparent.

To manage fraud, our system design takes a different approach that is oriented
toward control mechanisms and transaction analytics rather than counterparty
profiling. Because every transaction involves a regulated financial
intermediary that would presumably be bound by AML/KYC regulations, there is a
clear path to investigating every transaction effectively. Authorities would
be positioned to ensure that holders of accounts that take payments from
non-custodial wallets adhere to certain rules and restrictions, including but not
limited to tax monitoring. The records from such accounts, combined with the
auditable ledger entries generated by the DLT system, could enable real-time
collection of data concerning taxable income that could support reconciliation
and compliance efforts. Because all of the retail payments involving digital
currency would ultimately use the same ledger, identification of anomalous
behaviour, such as a merchant supplying an invalid destination account for
remittances from non-custodial wallets, would be more straightforward than in the
current system, and real-time automated compliance would be more readily
achievable. Such detection could even be done in real-time not only by
authorities but also by customers, thus reducing the likelihood that it would
occur in the first instance.

It is worth considering whether safely storing large amounts of physical cash
would be more or less costly than storing large amounts of digital currency.
In principle, digital currency can be stored cheaply online, although the
attack surface of online systems might have important weaknesses, and the
longevity of offline digital media has limits. Note that security safes are
generally priced as a function of the value, not the storage cost, of what is
inside. In addition, the use of token ``vintages'', perhaps arranged by year of
issue, can explicitly penalise the accumulation of large stashes of digital
currency in a manner that is hard to replicate with physical cash.  (Tokens
within one vintage would be fungible with each other, but not with tokens of
other vintages.)

It is also worth considering whether criminal organisations might exchange
private keys rather than entering transactions on the ledger as a way to avoid
interacting with MSBs. Our view is that sharing a private key is equivalent to
sharing the ability to spend money that can only be spent once, effectively
constituting a promise, otherwise as transferring possession in the case of a
non-custodial wallet. (Note that every token would have its own private key.)
Criminals can exchange promises by a variety of private or offline methods even
in the absence of a privacy-respecting payment system. At one level, it is
impossible to monitor or restrict such exchanges of promises, but at another
level, exchanges of this sort would require a high degree of a priori trust to
succeed, and we submit that transitive trust relationships would generally
degrade rapidly across successive transactions.  Meanwhile, attempts to spend
the same token twice can be easily detected, and potentially investigated, by
authorities at the time of the transaction. In our view, the utility derived
from the privacy preserving nature of a payment infrastructure warrants a
trade-off; however, the trade-off is substantially limited given the added
capability available to law enforcement and the mechanisms that may be
instituted, coupled with the fact that would there to be nefarious actors and
activities, those activities could take place in a variety of ways and media,
and they are not more effectively enabled by our system.

\subsection{Comparison to Alternative Approaches}
\label{ss:comparison}

Table~\ref{t:table} offers a
comparison of the main design features. The features of our design that
contrast with many of the prevailing CBDC design proposals include, but are not
limited to, the following:

\ifnum\x=0
  \end{paracol}
\fi
\begin{table}[t]
\widefigure
\begin{center}

\sf
\begin{tabular}{|L{9.3cm}|p{\ccol}p{\ccol}p{\ccol}p{\ccol}p{\ccol}p{\ccol}p{\ccol}p{\ccol}p{\ccol}|}\hline
& \rotatebox{90}{Goodell, Al-Nakib, Tasca}
& \rotatebox{90}{R3~\cite{r3-cbdc}}
& \rotatebox{90}{Bank of England~\cite{boe2020}}
& \rotatebox{90}{Sveriges Riksbank~\cite{riksbank}}
& \rotatebox{90}{Adrian and Mancini-Griffoli (IMF)~\cite{adrian2019}\,\,\,}
& \rotatebox{90}{Bordo and Levin~\cite{16}}
& \rotatebox{90}{ConsenSys~\cite{consensys}}
& \rotatebox{90}{Zhang ``Synthetic CBDC'' (IMF)~\cite{zhang2020}}
& \rotatebox{90}{Auer and B\"ohme (BIS)~\cite{auer2020a}}\\\hline
Can hold value outside an account                  & \CIRCLE & \Circle & \Circle & \Circle & \Circle & \Circle & \Circle & \Circle & \Circle \\
DLT system                             & \CIRCLE & \CIRCLE & \Circle & \CIRCLE & \Circle & \Circle & \CIRCLE & \Circle & \Circle \\
No central gatekeeper for transactions               & \CIRCLE & \CIRCLE & \Circle & \CIRCLE & \CIRCLE & \Circle & \CIRCLE & \CIRCLE & \Circle \\
Can be operated exclusively by private, independent actors     & \CIRCLE & \CIRCLE & \Circle & \CIRCLE & \CIRCLE & \Circle & \CIRCLE & \CIRCLE & \Circle \\
State manages issuance and destruction               & \CIRCLE & \CIRCLE & \CIRCLE & \CIRCLE & \CIRCLE & \CIRCLE & \CIRCLE & \Circle & \CIRCLE \\
Retail users do not hold accounts with the central bank       & \CIRCLE & \CIRCLE & \CIRCLE & \Circle & \CIRCLE & \Circle & \Circle & \CIRCLE & \CIRCLE \\
True privacy (in contrast to data protection)            & \CIRCLE & \Circle & \Circle & \Circle & \Circle & \Circle & \Circle & \Circle & \Circle \\
All transactions are on-ledger                   & \CIRCLE & \CIRCLE & \CIRCLE & \CIRCLE & \Circle & \CIRCLE & \CIRCLE & \Circle & \CIRCLE \\
All transactions require a regulated intermediary          & \CIRCLE & \CIRCLE & \CIRCLE & \Circle & \CIRCLE & \CIRCLE & \Circle & \CIRCLE & \CIRCLE \\
Intermediaries can include non-financial institutions        & \CIRCLE & \Circle & \Circle & \Circle & \Circle & \Circle & \CIRCLE & \Circle & \Circle \\
\hline\end{tabular}
\rm

\caption{\cz{Comparison of features among proposed retail digital currency architectures.}}

\label{t:table}
\vspace{-1em}
\end{center}
\end{table}
\ifnum\x=0
  \begin{paracol}{2}
  \linenumbers
  \switchcolumn
\fi

\begin{enumerate}

\item\cz{Retail users can hold digital assets outside accounts.} Most of the
existing proposals assume that digital assets would be always held by
intermediaries. In contrast, our proposal empowers retail users with the
ability to truly control the assets they hold and choose custodians, when
applicable, on their own terms.

\item\cz{No central bank accounts for individuals and non-financial
businesses.} In our view, requiring central bank accounts would introduce new
costs, weaknesses, and security vulnerabilities. It would result in the
central bank taking responsibility for actions commonly performed by the
private sector in many countries, and it would negate the benefits of using
tokens rather than accounts. A team led by Jes\'us Fern\'andez-Villaverde
observed that many proponents of CBDC, such as Bordo and Levin~\cite{16}, assume
that central banks would disintermediate commercial intermediaries and that in
many cases this possibility is touted as a benefit of
CBDC~\cite{fernandez2020}. However, their analysis formalises a trade-off
between avoiding bank runs and delivering optimal allocation of
capital~\cite{fernandez2020}, underscoring a key role of commercial banks in
bearing risk that, in our view, should not be undermined.

\item\cz{A purpose-built domestic, retail payment system.} The requirement to
support cross-border or wholesale payments is intentionally not included in our
design. Our proposal is designed specifically to meet the requirements for a
domestic, retail payment system, which we believe differ significantly from the
requirements for a cross-border or wholesale payment system.

\item\cz{True, verifiable privacy for retail users.} Data protection is not
the same as privacy, and our proposal does not rely upon third-party trust or
data protection for their transaction metadata. Some proposals include
``anonymity vouchers'' that would be usable for a limited time in
accounts-based digital currency systems~\cite{r3-cbdc,dgen}. We do not believe
that such approaches would be effective, not only because of the dangers
associated with reducing the anonymity set to specific intervals but also
because of the attacks on anonymity that will always be possible if value is to
be transferred from one regulated account directly to another.

\item\cz{No new digital identity systems.} Our system does not require any
special identity systems beyond those that are already used by MSBs and
private-sector banks. In particular, it does not require a system-wide
identity infrastructure of any kind, and it also explicitly allows individuals
to make payments from their non-custodial wallets without revealing their
identities.

\item\cz{No new real-time operational infrastructure managed by central
authorities.} Our proposed system can be operated exclusively by private,
independent actors without relying upon a central actor to operate any specific
part of the infrastructure. The distributed ledger makes it possible to assign
responsibility for most transactions to the MSBs, not the central bank. An MSB
is responsible for each transaction that it writes to the ledger, and the DLT
can be used to create a (potentially) immutable record binding every
transaction to the corresponding MSB that submitted it. We understand that the
central bank is not responsible for individual transactions.

\end{enumerate}

\section{Recommendations}

We believe that all the models proposed so far for CBDC fail to meet important
design criteria that have been summarised in Table~\ref{t:table}. In
particular, we have shown that other concurrent CBDC design proposals omit
certain design features that have an impact on critical areas of
welfare-generating characteristics, as well as governance and financial
implications. The proposal that we have articulated addresses these essential
requirements directly and does not compromise.

The following design features make our model unique. First, our proposal uses
a DLT-based settlement system that is overseen by State actors but operated
entirely by private, independent actors. Second, it aims to enhance the
welfare and safety of users by employing \textit{privacy by design} without
compromising the core risk analysis capacity in which policymakers would find
value.

In all cases, it is critical to separate the regulatory requirements for
identification (the `policy') from the underlying protocols and technology that
facilitate payments (the `mechanism'). Such separation must be seen as a
requirement for non-custodial wallets. The mechanism by which custodial retail
electronic payments are implemented enables surveillance as an artifact of the
custodial relationship. For owners of money to truly use it freely, they must
have a means of using money outside custodial relationships and without the
risk of profiling. To impose requirements upon non-custodial wallets that
essentially proscribe such uses would only serve to ensure that digital money
is never truly owned, as its users would be forced to accept a more limited set
of rights.\footnote{This paragraph also appears in a response to a recent
consultation by the US Financial Crimes Enforcement
Network~\cite{goodell2021}.}

\section{Conclusion}

With a guiding principle that it is not possible to trust something that is not
possible to verify, we have shown that it is feasible to design a digital
payment system that combines the most salient features of cash with the most
salient features of regulatory oversight. We have shown how privacy-enhancing
technology can protect users from profiling by allowing anonymous
counterparties, even if the transactions are not peer-to-peer and regulators
can observe every transaction. We have shown how distributed ledger technology
can avoid the costs and risks of centralised infrastructure operated by
governments or their contractors, while allowing the public to verify that the
system operates as advertised. We have concluded that it is both possible and
necessary to allow users to hold value outside of custodial relationships, and
we have shown how it is possible to implement an effective CBDC system while
allowing CBDC users to preserve their existing banking relationships and
avoiding the need for centralised accounts or identity management systems.

We hope that policymakers and business leaders will agree with us that
preserving the rights of individual persons is essential in the context of
e-commerce and electronic retail payments, particularly as such systems capture
an ever-increasing share of the retail economy. Current electronic retail
payment infrastructure exposes its users to risks including but not limited to
profiling, discrimination, and reduced autonomy. Our research shows that we
can choose a better future with a purpose-built, decentralised domestic retail
payment system that serves the public interest.

\section*{Acknowledgements}

We thank Professor Tomaso Aste for his continued support for our project, and
we thank Larry Wall of the Federal Reserve Bank of Atlanta, Robleh Ali of the
MIT Media Laboratory, and Erica Salinas of the Value Technology Foundation for
their valuable feedback. We also acknowledge the support of the Centre for
Blockchain Technologies at University College London, the Centre for Technology
and Global Affairs at the University of Oxford, and the Systemic Risk Centre at
the London School of Economics, and we specifically acknowledge the European
Commission for the FinTech project (H2020-ICT-2018-2 825215).

\sf

\vspace{1em}
\noindent All icons and clipart images are available at
publicdomainvectors.org, with the exception of the wallet icon, which is
available at vectorportal.com.

\end{document}